\documentclass[aps,pre,onecolumn,floatfix]{revtex4}

\usepackage{bm}
\usepackage{graphicx}
\usepackage{amssymb,amsfonts,amsmath}
\usepackage{epsf}
\usepackage{color}
\usepackage{hyperref}
\usepackage{subfigure}
\usepackage[export]{adjustbox}
\usepackage{epstopdf}
\usepackage{rotating}
\usepackage{soul}

\usepackage{mathrsfs}

\usepackage{upgreek}

\newcommand{\be}{\begin{equation}}
\newcommand{\ee}{\end{equation}}
\newcommand{\bea}{\begin{eqnarray}}
\newcommand{\eea}{\end{eqnarray}}

\begin{document}

\title{\bf Elastic and transport coefficients of the perfect hard-sphere crystal from the poles of the hydrodynamic spectral functions }

\author{Jo\"el Mabillard}
\email{Joel.Mabillard@ulb.be; \vfill\break ORCID: 0000-0001-6810-3709.}
\author{Pierre Gaspard}
\email{Gaspard.Pierre@ulb.be; \vfill\break ORCID: 0000-0003-3804-2110.}
\affiliation{Center for Nonlinear Phenomena and Complex Systems, Universit{\'e} Libre de Bruxelles (U.L.B.), Code Postal 231, Campus Plaine, B-1050 Brussels, Belgium}

\begin{abstract}
The elastic and transport coefficients of a perfect face-centered cubic crystal of hard spheres are computed from the poles of the dynamic structure factor and of the spectral functions of transverse momentum density fluctuations. For such  crystals, the relevant coefficients are the three isothermal elastic constants $(C_{11}^T,C_{12}^T,C_{44}^T)$, the heat conductivity $\kappa$, and the three viscosities $(\eta_{11},\eta_{12},\eta_{44})$ (in Voigt's notations),  which are directly computed using molecular dynamics simulations. The elastic and transport coefficients are then compared to the values of the same coefficients obtained with the method of Helfand moments, showing good agreement and providing strong support for the microscopic hydrodynamic theory of perfect crystals based on the local-equilibrium approach.
\end{abstract}

\maketitle

\section{Introduction}
\label{sec:intro}

Dissipative hydrodynamics is a general theory to describe the large-scale properties of materials and, in particular, the propagation and attenuation of sound waves in crystals~\cite{MPP72,F75,FC76,KDEP90,SE93}.  On the one hand, the elastic properties and the mass density determine the propagation speeds of the sound waves.  On the other hand, their attenuation is caused by energy dissipation due to transport properties such as the heat conductivities and the viscosities of the crystal.  The transport coefficients can be calculated from the microscopic dynamics of the atoms composing the system, using the Green-Kubo or the Einstein-Helfand formulas~\cite{S97,MG21,H22,H23}.  These formulas are derived by considering the linear response of the system to initial perturbations of large spatial scales in the hydrodynamic regime.  

Likewise, the response of the system to external perturbations like photon or neutron scattering of given frequency and wave vector can be expressed in terms of spectral functions such as the dynamic structure factor~\cite{vH54,G55,BF66}.  The spectral functions present resonances associated with the intrinsic collective modes of the material.  At complex frequencies, the resonances have precise locations given by the poles of the spectral functions and corresponding to the dispersion relations of the modes.  Their real part depends on the propagation speeds and their imaginary part on the transport properties, which determine the width of the resonances and, thus, the lifetime of the modes.  Accordingly, the resonances of the spectral functions and their underlying poles can be used to obtain the hydrodynamic properties of the material. 

In this paper, our purpose is to validate the predictions of the local-equilibrium approach to the hydrodynamics of crystals by carrying out for the hard-sphere crystal the same programme we achieved for the hard-sphere fluid in reference~\cite{MG23}.  With this aim, we directly compute the poles of the spectral functions by performing molecular dynamics simulations of the hard-sphere system.  At high densities, this system forms a face-centered cubic (fcc) crystal, for which we have previously calculated the elastic and transport coefficients using the method of Helfand moments in reference~\cite{MG23_primo}.  Here, we compare and test these results with the determination of the same coefficients using the direct computational method.  For this purpose, we use the analytical expressions of the spectral functions derived from the hydrodynamics of perfect cubic crystals in reference~\cite{MG23_secundo}.  In this previous work, the dispersion relations of the hydrodynamic modes of these crystals have been obtained in terms of the elastic and transport coefficients.  Accordingly, the values of these coefficients can be determined from the resonances of the spectral functions.

The hard-sphere crystal we consider is perfect, meaning that every lattice site is occupied by exactly one sphere, which implies the absence of the vacancy diffusion mode. Therefore, perfect crystals have only seven out of the eight hydrodynamic modes generated by the five fundamental conservation laws and the spontaneous symmetry breaking of continuous translations in the three spatial directions.  The seven hydrodynamic modes include the six longitudinal and transverse sound waves and the heat conduction mode.  They have characteristic speeds and diffusivities depending on the elastic and transport coefficients.  Here, our goal is thus to evaluate these coefficients from the poles of the spectral functions and to compare their values with those obtained using the Einstein-Helfand formulas in reference~\cite{MG23_primo}.  More generally, the present study aims at showing that the viscosities and the heat conductivities of crystals can be estimated by measuring the widths of the resonances, for instance, in the spectra of neutron inelastic scattering~\cite{SBR67,SROR72,TES78}.

The paper has the following plan. In section~\ref{sec:SF}, the spectral functions characterizing the fluctuations of mass and transverse momentum densities are introduced.  Their analytical expressions derived in reference~\cite{MG23_secundo} are recalled and their poles give the dispersion relations of the seven hydrodynamic modes in the large-system limit, where hydrodynamics should hold.  In section~\ref{sec:Num}, the spectral functions are directly computed with molecular dynamics simulations for the hard-sphere system and their poles are numerically located at complex frequencies by fitting rational functions.  The dispersion relations are thus obtained from the simulations of the hard-sphere dynamics.  In this way, the speeds and attenuation coefficients of the sound waves are computed for different directions of the wave vector in the limit of small magnitude to reach the hydrodynamic regime.  In section~\ref{sec:ECTC}, the speeds and attenuation coefficients for the different directions are combined together in order to evaluate the elastic and transport coefficients of the perfect hard-sphere crystal, which can thus be compared to their values calculated with the method of Helfand moments in reference~\cite{MG23_primo}.  Section~\ref{sec:concl} presents the conclusion and the perspectives.

{\it Notations.} The Cartesian spatial coordinates are denoted by Latin indices such as $a = x, y, z$. Einstein's convention of summation over repeated indices is adopted unless explicitly stated.  $k_{\rm B}$ is Boltzmann's constant and ${\rm i}=\sqrt{-1}$.

\section{Spectral functions and their poles}
\label{sec:SF}

In this section, we introduce the spectral functions needed for the evaluation of the elastic and transport coefficients in one-component perfect crystals with cubic symmetry, namely the three isothermal elastic constants $(C_{11}^T,C_{12}^T,C_{44}^T)$, the heat conductivity $\kappa$, and the three viscosities $(\eta_{11},\eta_{12},\eta_{44})$ (in Voigt's notations). For this purpose, the spectral functions to consider are the dynamic structure factor and the spectral functions of transverse momentum density fluctuations, which have been calculated in reference~\cite{MG23_secundo} from the linearized equations of dissipative hydrodynamics~\cite{MG21}. We also explain how the poles of the spectral functions and the dispersion relations are related to the elastic and transport coefficients.


\subsection{Microscopic dynamics and fluctuating densities} 
\label{subsec:MDFD}

As in references~\cite{MG23_primo,MG23_secundo}, we consider a system of $N$ identical particles of mass $m$, positions ${\bf r}_i$, and momenta ${\bf p}_i$ with the labels $1\leq i \leq N$.  These particles move in a cubic spatial domain of sides $L$ and volume $V=L^3$ with periodic boundary conditions.  In the phase space  of the variables $\Gamma=({\bf r}_i,{\bf p}_i)_{i=1}^N\in{\mathbb R}^{6N}$, the time evolution is generated by the equations of motion of the microscopic dynamics, which conserves the total energy $E$ given by the Hamiltonian function $H(\Gamma)$ and the total momentum ${\bf P}=\sum_{i=1}^N{\bf p}_i$.  The dynamics also satisfies Liouville's theorem and the property of microreversibility.  Statistics is carried out using the $(N,V,E)$-ensemble and setting the total momentum to ${\bf P}=0$, which defines the statistical average $\langle\cdot\rangle_{\rm eq}$ with respect to equilibrium.

The microscopic observables of interest are the mass density $\hat\rho({\bf r},t)\equiv m\sum_{i=1}^N\delta[{\bf r}-{\bf r}_i(t)]$ and the momentum density $\hat{g}^a({\bf r},t)\equiv \sum_{i=1}^N p_i^a(t)\, \delta[{\bf r}-{\bf r}_i(t)]$, which are fields fluctuating in time $t$ and in space ${\bf r}$.  Their Fourier transforms from the space ${\bf r}$ to the  reciprocal space of wave vectors ${\bf q}$ are respectively given by
\begin{align}
\label{eq:rho(q)+g(q)}
\hat{\rho}({\bf q},t) = m \sum_{i=1}^N {\rm e}^{{\rm i} {\bf q}\cdot{\bf r}_i(t)}
\qquad\mbox{and}\qquad
\hat g^a({\bf q},t) = \sum_{i=1}^N p_i^a(t) \, {\rm e}^{{\mathrm i}{\bf q}\cdot{\bf r}_i(t)} \, .
\end{align}
Because of the periodic boundary conditions on the cubic domain $[0,L[^3$, the wave vector takes the discrete values ${\bf q} = (2\pi/L)\left(n_x{\bf e}_x+n_y{\bf e}_y+n_z{\bf e}_z \right)$ with $(n_x,n_y,n_z)\in {\mathbb Z}^3$ in the Cartesian basis of unit vectors $\{{\bf e}_x,{\bf e}_y,{\bf e}_z\}$.

Another orthonormal basis $\{{\bf e}_{\rm l},{\bf e}_{{\rm t}_1},{\bf e}_{{\rm t}_2}\}$ can be introduced, where ${\bf e}_{\rm l}\equiv{\bf q}/q$ with $q=\Vert{\bf q}\Vert$ is oriented in the direction of the wave vector $\bf q$ and ${\bf e}_{{\rm t}_k}$ with $k=1,2$ are oriented in the two transverse directions.  Accordingly, the Fourier transform of the momentum density, which is vectorial, can be decomposed into the corresponding components $\hat g_\sigma({\bf q},t) =e^a_\sigma\,\hat g^a({\bf q},t)$ with $\sigma\in\{{\rm l},{\rm t}_1,{\rm t}_2\}$, which are longitudinal and transverse with respect to the wave vector $\bf q$.

At equilibrium in the crystalline phase, the mean mass density is stationary and has the periodicity of the crystal lattice, $\langle\hat\rho({\bf r})\rangle_{\rm eq}=\sum_{\bf G} \rho_{{\rm eq},{\bf G}}\, {\rm e}^{-{\rm i}{\bf G}\cdot{\bf r}}$, where  the sum extends over the reciprocal lattice vectors $\bf G$.  In order to probe the hydrodynamic regime, we have to consider wave vectors $\bf q$ that are significantly smaller in magnitude than the smallest non-zero reciprocal lattice vectors $\bf G$, meaning that the hydrodynamic modes have a wavelength $\lambda=2\pi/q$ much larger than the size of the  primitive lattice cell.  Therefore, the deviations of the fluctuating modes with respect to equilibrium are given by $\delta\hat\rho({\bf q},t)\equiv\hat\rho({\bf q},t)-\langle\hat\rho({\bf q})\rangle_{\rm eq}=\hat\rho({\bf q},t)$, since $\langle\hat\rho({\bf q})\rangle_{\rm eq}=0$ for $0<\Vert{\bf q}\Vert\ll\Vert{\bf G}\Vert$ in the hydrodynamic regime, and $\delta\hat{g}_\sigma({\bf q},t)\equiv\hat{g}_\sigma({\bf q},t)-\langle\hat{g}_\sigma({\bf q})\rangle_{\rm eq}=\hat{g}_\sigma({\bf q},t)$, because $\langle\hat{g}_\sigma({\bf q})\rangle_{\rm eq}=0$.

The spacetime fluctuations of the Fourier modes can be characterized by their time-dependent correlation functions.  With the principle of regression of fluctuations~\cite{O31b}, analytical expressions can be derived for the correlation functions by solving the linearized equations for the macroscopic dissipative hydrodynamics of crystals \cite{MG23_secundo}.
For the perfect hard-sphere crystal we here consider, the lattice is fcc.  In such crystals, the set of equations splits into decoupled equations for the longitudinal and transverse components of the Fourier modes if the wave vector $\bf q$ is oriented into one of the special directions given in table~\ref{Tab:LTCoeffs} with the corresponding stress-strain and viscosity coefficients.


\begin{table}[h!]
  \begin{tabular}{c @{\hskip 1cm} c @{\hskip 1cm} c @{\hskip 1cm} c  }
    \hline\hline
      Direction&   $[100]$     &    $[110]$  &  $[111]$   	 \\
    \hline  
    ${\bf e}_{\rm l}$ & ${\bf e}_x$  	& $({{\bf e}_x+{\bf e}_y})/{\sqrt{2}}$ 	& $({{\bf e}_x+{\bf e}_y+{\bf e}_z})/{\sqrt{3}}$	\\
    $ {\bf e}_{{\rm t}_1}$		& $  {\bf e}_y$		& $({{\bf e}_x-{\bf e}_y})/{\sqrt{2}}$ & $ ({{\bf e}_x-{\bf e}_y})/{\sqrt{2}}$ \\
        $ {\bf e}_{{\rm t}_2}$		& $  {\bf e}_z$		& $  {\bf e}_z$ & $({{\bf e}_x+{\bf e}_y-2{\bf e}_z})/{\sqrt{6}}$ \\
            $B^T_{\rm l} $		& $B^T_{11}$		& $({B^T_{11}+B^T_{12}+2B^T_{44}})/{2}$ & $ ({B^T_{11}+2B^T_{12}+4B^T_{44}})/{3}$ \\
                $ B^T_{{\rm t}_1}$		& $B^T_{44}$		& $({B^T_{11}-B^T_{12}})/{2}$ & $ ({B^T_{11}-B^T_{12}+B^T_{44}})/{3}$ \\
                $ B^T_{{\rm t}_2}$		& $B^T_{44}$		& $B^T_{44}$ & $ ({B^T_{11}-B^T_{12}+B^T_{44}})/{3}$ \\
            $\eta_{\rm l} $		& $ \eta_{11}$		& $({\eta_{11}+\eta_{12}+2\eta_{44}})/{2}$ & $({\eta_{11}+2\eta_{12}+4\eta_{44}})/{3}$ \\
                $ \eta_{{\rm t}_1}$		& $\eta_{44}$		& $({\eta_{11}-\eta_{12}})/{2}$ & $({\eta_{11}-\eta_{12}+\eta_{44}})/{3}$ \\
                $ \eta_{{\rm t}_2}$		& $\eta_{44}$		& $\eta_{44}$ & $({\eta_{11}-\eta_{12}+\eta_{44}})/{3}$ \\                
    \hline\hline
  \end{tabular}
  \caption{The stress-strain coefficients $B^T_{\sigma}$ and the viscosity coefficients $\eta_{\sigma}$ in the longitudinal and transverse directions ${\bf e}_\sigma$ with $\sigma\in\{{\rm l},{\rm t}_1,{\rm t}_2\}$ for the wave vector ${\bf q}$ oriented in the directions $[100]$, $[110]$, and $[111]$ of the cubic crystal, as expressed using Voigt's notations. }\label{Tab:LTCoeffs}
\end{table}

Next, the frequency content of the time-dependent correlation functions can be obtained by calculating their spectral functions, which depend on the wave number $q$ and the frequency $\omega$.  The spectral functions have poles at complex frequencies, giving the dispersion relations of the hydrodynamic modes and allowing us to directly evaluate the elastic and transport properties by molecular dynamics simulations.

The following subsections~\ref{subsec:DSF} and~\ref{subsec:TSMF} present these functions associated with the mass and momentum densities, respectively.

\subsection{Dynamic structure factor} 
\label{subsec:DSF}

The time-dependent correlation function of the mass density fluctuations $F({\bf q},t)\equiv\langle\delta\hat\rho({\bf q},t)\, \delta\hat\rho^*({\bf q},0)\rangle_{\rm eq}/(Nm^2)$ is known as the {\it intermediate scattering function}~\cite{vH54,G55}.
Its Fourier transform from time to frequency defines the {\it dynamic structure factor} $S({\bf q},\omega)\equiv\int_{-\infty}^{+\infty} F({\bf q},t)\, {\rm e}^{-{\rm i}\omega t}\, {\rm d}t$.  

If the wave vector ${\bf q}$ is oriented in the directions $[100]$, $[110]$, and $[111]$, the dynamic structure factor for a perfect crystal of hard spheres has the following analytical form~\cite{MG23_secundo},
\begin{align}
\frac{S({q},\omega)}{S({q})} & = \frac{1}{1+(\gamma-1)\frac{B_T}{ B_{\rm l}^T}}\left\{(\gamma-1)\frac{B_T}{ B_{\rm l}^T} \frac{2 \chi q^2}{\omega^2+\left(\chi q^2\right)^2}+\frac{\Gamma_{\rm l} q^2}{(\omega+c_{\rm l}q)^2+(\Gamma_{\rm l} q^2)^2}+\frac{\Gamma_{\rm l} q^2}{(\omega-c_{\rm l}q)^2+(\Gamma_{\rm l} q^2)^2}\right.\notag\\
&\left.+\frac{3\Gamma_{\rm l}-D_v}{c_{\rm l}}q\left[\frac{\omega+c_{\rm l}q}{(\omega+c_{\rm l}q)^2+(\Gamma_{\rm l} q^2)^2}-\frac{\omega-c_{\rm l}q}{(\omega-c_{\rm l}q)^2+(\Gamma_{\rm l} q^2)^2}\right]\right\} ,\label{eq:DSF_RBP}
\end{align}
where $q=\Vert{\bf q}\Vert$, $S(q)=F(q,0)$ is the {\it static structure factor}, $\gamma$ the specific heat ratio, $B_T$ the isothermal bulk modulus, and $B_{\rm l}^T$ a linear combination of the isothermal stress-strain coefficients, which depends on the direction of  ${\bf q}$, as given in table~\ref{Tab:LTCoeffs}.  The coefficient $\chi$ is defined as
\begin{align}
\label{eq:chi}
\chi\equiv\frac{\gamma D_T}{1+(\gamma-1)\frac{B_T}{B_{\rm l}^T}} \, ,
\end{align}
where $D_T\equiv\kappa/(\gamma \rho c_v)$ is the thermal diffusivity, $\kappa$ being the heat conductivity, $c_v$ the specific heat capacity at constant specific volume, and $\rho$ the spatially averaged equilibrium mass density.  The speed of the longitudinal sound waves is expressed as
\begin{align}
\label{eq:csl}
c_{\rm l} \equiv\sqrt{\frac{B_{\rm l}^T+(\gamma-1)B_T}{\rho}}
\end{align}
and their acoustic attenuation coefficient as
\begin{align}
\label{eq:Gaml}
\Gamma_{\rm l}\equiv\frac{1}{2}\left(D_v  + \frac{\gamma D_T}{1+\frac{1}{\gamma-1}\frac{B_{\rm l}^T}{B_T}} \right) ,
\end{align}
where $D_v\equiv \eta_{\rm l}/\rho$ is the longitudinal kinematic viscosity and $\eta_{\rm l}$ a linear combination of the viscosities which depends on the direction of  ${\bf q}$, as given in table~\ref{Tab:LTCoeffs}.

The poles of the dynamic structure factor $S({q},\omega)$ are located at the  complex frequencies
\begin{align}
\label{eq:DR_DSF}
\omega_0(q) &= {\rm i}\, \chi \, q^2+\cdots\,, && \omega_{{\rm l}\pm}(q) = \pm  c_{\rm l}\, q + {\rm i}\, \Gamma_{\rm l} \, q^2+\cdots \, ,
\end{align} 
and their complex conjugates  $\omega_0^*(q)$ and $\omega^*_{\rm l\pm}(q)$. From the dispersion relations~\eqref{eq:DR_DSF}, we identify the heat mode, which is purely diffusive with a damping determined by the coefficient~\eqref{eq:chi}, and the pair of longitudinal sound modes propagating with the speeds $\pm  c_{\rm l}$. The damping of these sound modes is proportional to the longitudinal acoustic attenuation coefficient~\eqref{eq:Gaml}.


\subsection{Spectral functions of transverse momentum density fluctuations}
\label{subsec:TSMF}

The momentum density fluctuations are characterized by the time-dependent correlation functions $C_\sigma({\bf q},t)\equiv\langle\delta\hat{g}_\sigma({\bf q},t)\, \delta\hat{g}_\sigma^*({\bf q},0)\rangle_{\rm eq}/(Nm^2)$ and the corresponding spectral functions $J_\sigma({\bf q},\omega)\equiv\int_{-\infty}^{+\infty} C_\sigma({\bf q},t)\, {\rm e}^{-{\rm i}\omega t}\, {\rm d}t$ for $\sigma\in\{{\rm l},{\rm t}_1,{\rm t}_2\}$.  In the two directions ${\rm t}_1$ and ${\rm t}_2$ that are transverse to the wave vector $\bf q$, when oriented in the directions $[100]$, $[110]$, and $[111]$, the linear hydrodynamics of crystals shows that the corresponding spectral functions can be expressed as~\cite{MG23_secundo}
\begin{align}
\frac{J_{{\rm t}_k}({q},\omega)}{C_{{\rm t}_k}({q},0)} & = \frac{2\, \eta_{{\rm t}_k}q^2\omega^2/\rho}{\left(\omega^2-B_{{\rm t}_k}^Tq^2/\rho\right)^2+\left(\eta_{{\rm t}_k}q^2\omega/\rho\right)^2}
\label{eq:JtoCt}
\end{align}
with $k=1,2$ and $q=\Vert{\bf q}\Vert$, where $\eta_{{\rm t}_k}$ and $B_{{\rm t}_k}^T$ are linear combinations of the viscosities and the isothermal stress-strain coefficients, depending on the direction of  ${\bf q}$, as given in table~\ref{Tab:LTCoeffs}.

The poles of  $J_{{\rm t}_k}({q},\omega)$ are located at the  complex frequencies
\begin{align}
\omega_{{\rm t}_k\pm}(q) & = \pm c_{{\rm t}_k}q+ {\rm i}\, \Gamma_{{\rm t}_k} q^2 + \cdots
 \qquad\mbox{  for}\qquad k=1,2 
\label{eq:poles_Jtqw}
\end{align}
and their complex conjugates $\omega_{{\rm t}_k\pm}^*(q)$, where
\begin{align}
\label{eq:cst}
c_{{\rm t}_k}&\equiv \sqrt{\frac{B_{{\rm t}_k}^T}{\rho}}
\end{align} 
are the speeds of the transverse sound waves and 
\begin{align}
\label{eq:Gamt}
\Gamma_{{\rm t}_k}&\equiv \frac{\eta_{{\rm t}_k}}{2\rho}
\end{align} 
the transverse acoustic attenuation coefficients. From the dispersion relations~\eqref{eq:poles_Jtqw},  we identify the two pairs of transverse sound modes propagating with the speeds $\pm c_{{\rm t}_k}$ and damped by the transverse acoustic attenuation coefficients~\eqref{eq:Gamt}.

Therefore, we have identified the seven hydrodynamic modes of the perfect crystal from the dispersion relations~\eqref{eq:DR_DSF} and~\eqref{eq:poles_Jtqw}, i.e., from the poles of the spectral functions.


\section{Numerical location of the poles of the spectral functions}
\label{sec:Num}

In this section, the poles of the spectral functions are obtained with molecular dynamics simulations of the hard-sphere dynamics. The time-dependent correlation functions are directly computed with the simulations and the spectral functions are  obtained with  numerical Fourier transforms from time to frequency. The poles are obtained from the roots of rational functions fitted to the spectral functions obtained from the simulations. Finally, the asymptotic values of the poles at $q=0$ are obtained with a linear least square regression over the values of the poles at finite $q$.

The results   will be presented in terms of dimensionless quantities denoted with an asterisk as subscript, after their rescaling in terms of the mass $m$ and the diameter $d$ of the hard spheres and the thermal energy $k_{\rm B}T$.  The particle density, wave number, frequency, isothermal bulk modulus, stress-strain coefficients, elastic coefficients, sound speeds, diffusivities, spectral functions, and transport coefficients  will respectively be given in terms of the corresponding dimensionless quantities by
\begin{align}
& n = \frac{n_*}{d^3} \, , \qquad q = \frac{q_*}{d}\, , \qquad \omega = \frac{\omega_*}{d}\, \sqrt{\frac{k_{\rm B}T}{m}} \, , \\
& B_T = B_{T*} \, \frac{k_{\rm B}T}{d^3} \, , \qquad B^T_\sigma = B^T_{\sigma*} \, \frac{k_{\rm B}T}{d^3} \, , \qquad C^T_\mu = C^T_{\mu*}  \, \frac{k_{\rm B}T}{d^3} \, ,\\
& c_\sigma = c_{\sigma *} \sqrt{\frac{k_{\rm B}T}{m}} \, , \qquad 
\Gamma_\sigma= \Gamma_{\sigma *} \, d \, \sqrt{\frac{k_{\rm B}T}{m}} \, , \qquad
\chi = \chi_* \, d \, \sqrt{\frac{k_{\rm B}T}{m}} \, , \\
& \frac{S(q,\omega)}{S(q)} = \left[ \frac{S(q,\omega)}{S(q)} \right]_* \, d \, \sqrt{\frac{m}{k_{\rm B}T}} \, ,
\qquad\frac{J_\sigma(q,\omega)}{C_\sigma(q,0)} = \left[ \frac{J_\sigma(q,\omega)}{C_\sigma(q,0)} \right]_* \, d \, \sqrt{\frac{m}{k_{\rm B}T}}\,,\\
& \eta_\mu = \eta_{\mu *} \, \frac{m}{d^2}\sqrt{\frac{k_{\rm B}T}{m}} \, , \qquad \mbox{and}\qquad
\kappa = \kappa_* \, \frac{k_{\rm B}}{d^2}\sqrt{\frac{k_{\rm B}T}{m}} \, , 
\end{align}
where $\sigma\in\{{\rm l},{\rm t}_1,{\rm t}_2\}$ and $\mu\in\{11,12,44\}$.


\subsection{Computation of the spectral functions}
\label{subsec:MDHS}

The intermediate scattering function $F({\bf q},t)$ and the correlation functions $C_{{\rm t}_k}({\bf q},t)$, where $k=1,2$, of transverse momentum density fluctuations are numerically computed by simulating the dynamics of the hard-sphere system using an event-driven algorithm, as explained in references~\cite{MG23,MG23_primo,MG23_secundo}.  The hard-sphere system forms a fcc crystal for   densities in the range $1.037\pm 0.003 \leq n_* < \sqrt{2}$.  Here, the two densities $n_*=1.037$ and $n_*=1.3$ have been selected for the computations.  Statistics is carried out over $10^4$ trajectories for $F({\bf q},t)$ and $5\times10^3$ trajectories for $C_{{\rm t}_k}({\bf q},t)$ with discrete time steps $\Delta t_*=0.01$.  The number of time steps $n_{\rm steps}$ is taken in order for the correlation functions to have decayed  to a small enough value over the time interval $n_{\rm steps}\Delta t$ of the trajectory.  The transverse correlation functions have to be simulated over a longer time interval than the intermediate scattering functions, since their damping is typically smaller. We have therefore used lesser statistics for the transverse functions, to reduce the computational cost.

We compute the intermediate scattering function and the correlation functions of transverse momentum density fluctuations for the first three wave vectors ${\bf q}= (2\pi/L)(n_x{\bf e}_x + n_y{\bf e}_y + n_z{\bf e}_z)$ in each direction of table~\ref{Tab:LTCoeffs}, respectively taking $n_x\in\{1,2,3\}$ and $n_y=n_z=0$ in the direction $[100]$, $n_x=n_y\in\{1,2,3\}$ and $n_z=0$ in $[110]$, and $n_x=n_y=n_z\in\{1,2,3\}$ in $[111]$.  Next, the spectral functions $S({q},\omega)$ and $J_{{\rm t}_k}({q},\omega)$ with $q=\Vert{\bf q}\Vert$ are obtained by numerical Fourier transforms from time to frequency. 

We have chosen to perform the simulations with a fixed number of particles, $N=2048$. This choice is motivated by the following reasons. On the one hand,  in order to be in the hydrodynamic regime, one typically needs to consider wave vectors ${\bf q}$ with a magnitude that is small enough. The wave vector  ${\bf q}$ is however constrained by the periodic boundary conditions on the spatial cubic domain $[0,L[^3$.  Since $L~\sim N^{1/3}$, increasing the number of particles allows reaching smaller values of $q$. Moreover, the elastic and transport coefficients, that we aim to obtain eventually, have a dependence on $1/N^{\alpha}$, where $\alpha=1$ for all the quantities but the heat conductivity where $\alpha=2/3$~\cite{MG23_primo}. Large values of $N$ are thus required. On the other hand, the computation of the correlation functions requires looking at time intervals that are long enough, such that the functions have decayed over the interval. In molecular dynamics simulations of hard spheres, the particles evolve in free flights, interrupted by binary elastic collisions. It is therefore the number of collisions that plays the role of time in the simulation, instead of the physical time. When increasing the number of particles, the number of collisions increases as well, so does the computational cost. We have found that $N=2048$ is a good balance between reaching a small enough $q$ in all the directions, and keeping a reasonable computational cost. This choice is sufficient to show the agreement between the method of   Helfand moments and the poles, and to obtain the elastic and  transport coefficients from the poles. 

 The spectral functions considered in this analysis are shown in figures~\ref{Fig:CF100-1.037}-\ref{Fig:CF111-1.3}  for the three aforementioned values of the wave number $q$ in each one of the directions $[100]$, $[110]$, and $[111]$, and the two densities $n_*=1.037$ and $n_*=1.3$. The same figures show a comparison with the spectral functions obtained analytically from hydrodynamics, according to equations~\eqref{eq:DSF_RBP} and~\eqref{eq:JtoCt}, using the parameters calculated with the method of Helfand moments~\cite{MG23_secundo,MG23_primo}. In particular, the quantities $c_{{\rm l}}$, $\Gamma_{{\rm l}}$, $\chi$, $c_{{\rm t}_k}$, and  $\Gamma_{{\rm t}_k}$ are obtained  from the hydrostatic pressure, the isothermal  stress-strain coefficients  $(B^T_{11},B^T_{12},B^T_{44})$, and the transport coefficients, namely the three viscosities  $(\eta_{11},\eta_{12},\eta_{44})$, and the heat conductivity~$\kappa$, which are directly calculated using the Einstein-Helfand formulas and the Helfand moments~\cite{MG21}. The agreement between the simulations and hydrodynamics is always excellent for the smallest values of $q$ in each direction, as already shown in reference~\cite{MG23_secundo}. At larger $q$, the difference between simulations and hydrodynamics is explained by the fact that we are moving away from the hydrodynamic regime.

Figures~\ref{Fig:CF100-1.037}-\ref{Fig:CF111-1.3} also show that the resonances of the spectral functions are more and more separated and they become broader and broader as the wave number $q$ increases.  The reason is that the separation between the resonances goes as $c_\sigma q$, while their widths increase as $\chi q^2$ and $\Gamma_\sigma q^2$, i.e., faster than their separation for increasing values of the wave number $q$.  The broader the resonance, the shorter the lifetime of the corresponding mode.  For the dynamic structure factor $S(q,\omega)$, this trend even leads to the overlap of the resonances for the largest values of $q$, as observed in the top panels of figures~\ref{Fig:CF100-1.037}-\ref{Fig:CF111-1.3}.  This can be explained by the presence of the central resonance due to the heat mode and the fact that the diffusivities $\chi$ and $\Gamma_{\rm l}$ are always significantly larger than $\Gamma_{{\rm t}_{1,2}}$, as seen in tables~\ref{Tab:CC1.037_100}-\ref{Tab:CC1.3_111}.

Furthermore, the comparison between figures~\ref{Fig:CF100-1.037}-\ref{Fig:CF111-1.037} for the density $n_*=1.037$ and figures~\ref{Fig:CF100-1.3}-\ref{Fig:CF111-1.3} for $n_*=1.3$ shows that the resonances of the spectral functions are more separated and broader for higher densities.  This behavior is consistent with the increase of the speeds $c_\sigma$ and the diffusivities $\chi$ ad $\Gamma_\sigma$ with the density $n_*$.

 In the following, we are thus interested in finding the locations of the poles for different values of the wave number~$q$ using the spectral functions computed with molecular dynamics simulations.

\subsection{Location of the poles of the spectral functions} 

For each spectral function obtained from the simulations, the poles are obtained as follows. First, the rational functions  
\begin{align}\label{eq:rational_fn}
R(\omega) = \frac{\sum_{k=0}^N b_k\, \omega^k}{1+\sum_{k=1}^M a_k\, \omega^k}
\end{align}
 are fitted to the spectral function with a nonlinear least square method, giving the coefficients $\{a_k\}$ and $\{b_k\}$. Since the spectral functions are vanishing in the limit of large frequencies, the  degree $M$ of the polynomial  in the denominator is always chosen to be larger than the degree $N$ in the numerator. The degrees of the polynomials in the denominator and numerator are thus taken as $6\leq M  \leq 24$ and $1\leq N \leq M -2$ for the dynamic structure factor  and $4 \leq M \leq 19$ and $0 \leq N \leq M -2$ for the spectral functions of transverse momentum density fluctuations. Since $J_{{\rm t}_k}({q},\omega)$ has two resonances and $S({q},\omega)$ three,  smaller values of $M$ can be taken for the rational functions fitted to the former  than the latter.
 
\begin{figure}[h!] \centering
    \begin{minipage}{0.5\textwidth}
        \centering
        \includegraphics[width=1.\textwidth]{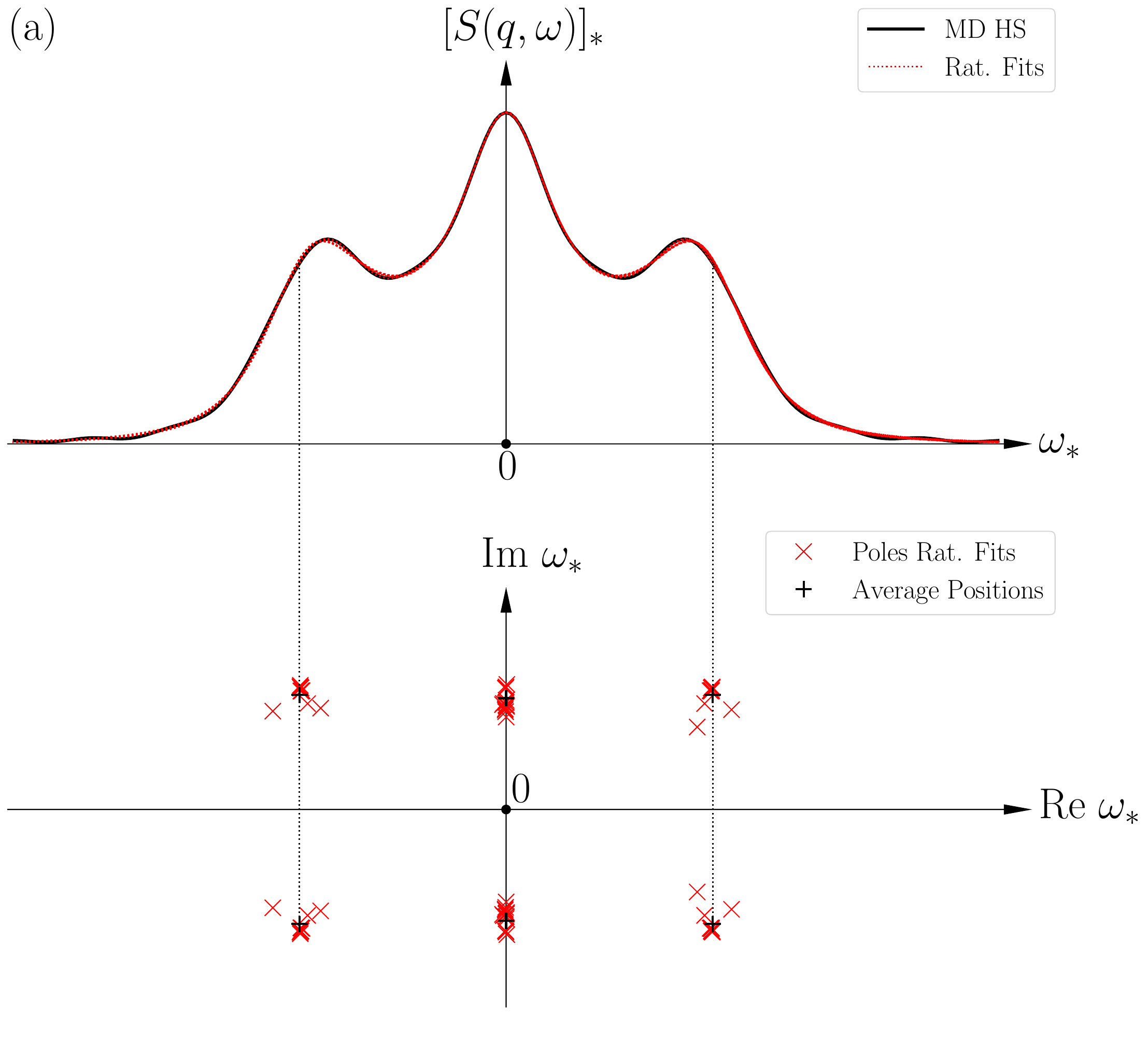} 
    \end{minipage}\hfill
     \begin{minipage}{0.5\textwidth}
        \centering
        \includegraphics[width=1.\textwidth]{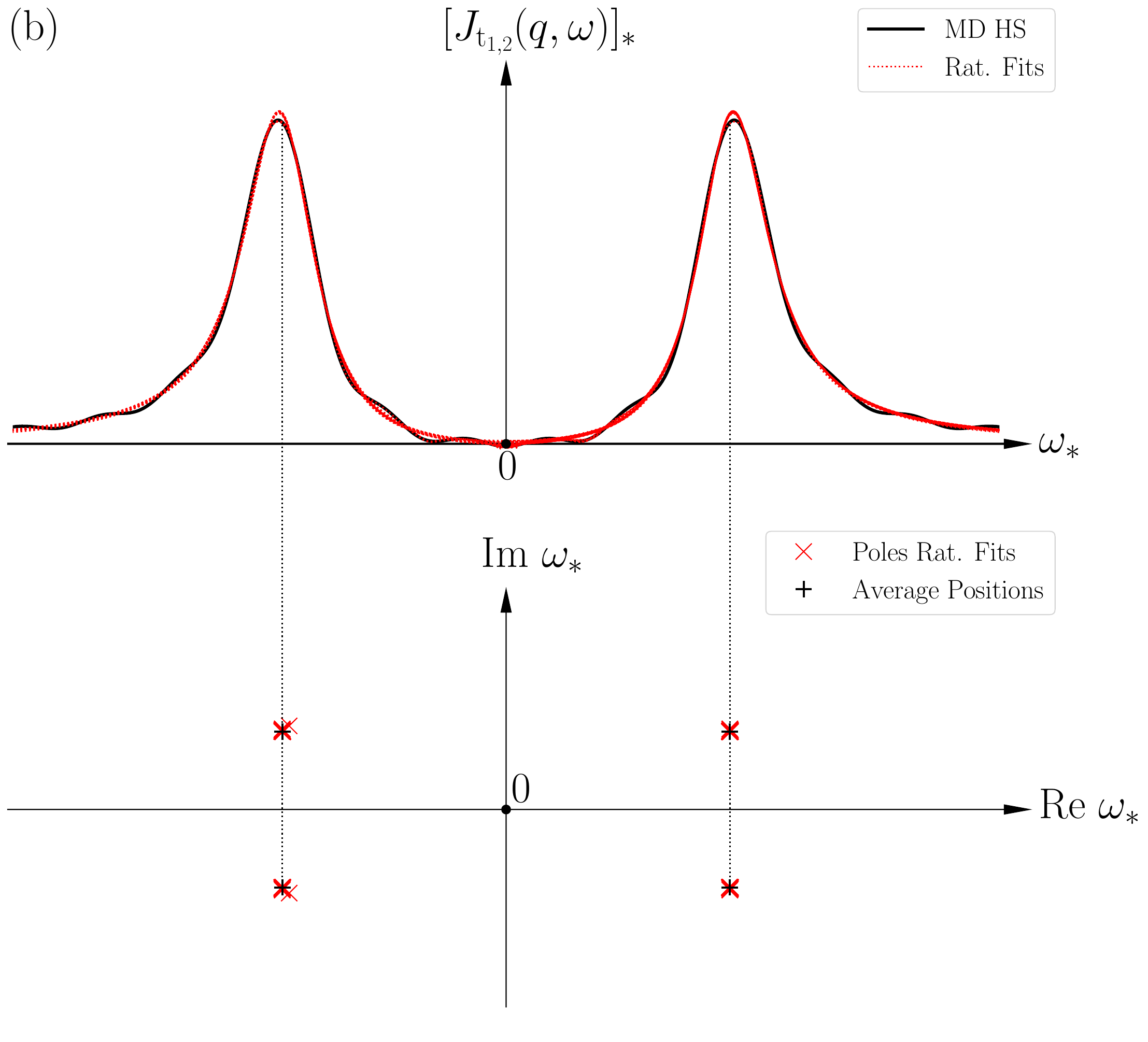} 
    \end{minipage}
\caption[] {Schematic representation of the method to locate  the poles.  The dynamic structure factor $S(q,\omega)$ and the  spectral functions of transverse momentum density fluctuations $J_{{\rm t}_k}(q,\omega)$  are obtained versus frequency $\omega$ from molecular dynamics simulations of the hard-sphere dynamics  (solid lines). The rational functions~\eqref{eq:rational_fn} with different values for the degrees $M$ and $N$ are fitted to the spectral functions (dotted  lines). The poles of the rational functions are obtained from the roots of the denominator (crosses) and the removal of spurious roots. Their mean locations (pluses) are obtained by taking the averages of the positions of the poles of the rational functions in a cluster. The poles are not always exactly located below the maximum of a peak, due to the  presence of nearby peaks. This shift is mostly seen for the Brillouin doublet on the left-hand side in the panel (a), and is mainly due to the central peak. The figure shows the spectral functions at density $n_*=1.3$, with ${\bf q}$ in the [111] direction and magnitude  $q_*=0.94$ under similar conditions as in the left panel of figure~\ref{Fig:CF111-1.3} corresponding to data in the third line of table~\ref{Tab:CC1.3_111}. }\label{Fig:SFPoles}
\end{figure}

The poles of the fitted rational functions are given by the roots of the denominator, after the removal of spurious  roots. Plotted in the complex plane of the frequencies, the poles of all the fitted rational functions appear to be grouped into clusters, as illustrated in figure~\ref{Fig:SFPoles}. We do not impose any even symmetry on the rational functions~\eqref{eq:rational_fn}, which explains the asymmetry observed in the spatial distribution of the poles of the fitted rational functions. The poles of the spectral function are finally determined by averaging over the locations of the poles of the  different fitted rational functions in the same cluster.  The error on the location of the poles is given by the standard  deviation.  As already observed in reference~\cite{MG23_secundo}, the poles are not always exactly located below the maximum of a peak. This shift stems from the presence of nearby peaks, creating an asymmetry,   which here mainly affects the Brillouin doublet of the dynamic structure factor.

\subsection{Asymptotic values of the poles}

The poles of the spectral functions are located at the complex frequencies $\omega_r(q)$, which depend on the wave number~$q$.  The leading terms expected for this dependence are given by equations~\eqref{eq:DR_DSF} and~\eqref{eq:poles_Jtqw} for the dispersion relations of the hydrodynamic modes.  In general, we can define $q$-dependent speeds of sound and diffusivities according to
\begin{align}\label{eq:q-coefficients}
c_\sigma(q) \equiv {\rm Re}[\omega_{\sigma +}(q)]/q \, , \qquad
\Gamma_\sigma(q) \equiv {\rm Im}[\omega_{\sigma\pm}(q)]/q^2 \, ,  \qquad\mbox{and} \qquad
\chi(q) \equiv {\rm Im}[\omega_0(q)]/q^2 \, ,
\end{align}
where $\sigma\in\{{\rm l},{\rm t}_1,{\rm t}_2\}$.
The extrapolation of these $q$-dependent values to $q=0$ should agree with the values given by the formulas~\eqref{eq:chi}, \eqref{eq:csl}, \eqref{eq:Gaml}, \eqref{eq:cst}, and~\eqref{eq:Gamt} in the hydrodynamic regime.

For the wave vector ${\bf q}$ oriented in a given direction, the determination of the poles for the same spectral function is repeated for different values of  the wave number $q$. In this work, we  consider the first three values in each direction. The extrapolation to $q=0$ is obtained with a least square linear regression over the dependence on $q^2$ for the quantities of interest. Details on the extrapolation and the estimation of the error are given in appendix~\ref{app:extq0}.

For the density $n_*=1.037$, the results are reported in tables~\ref{Tab:CC1.037_100},~\ref{Tab:CC1.037_110} and~\ref{Tab:CC1.037_111} for the directions [100],  [110], and  [111] respectively, and  shown in figure~\ref{Fig:LR_rho1.037}.  For $n_*=1.3$, the results are reported in tables~\ref{Tab:CC1.3_100},~\ref{Tab:CC1.3_110}, and~\ref{Tab:CC1.3_111} for the  same directions respectively, and shown in figure~\ref{Fig:LR_rho1.3}. The results are compared to the values obtained for the coefficients~\eqref{eq:chi}, \eqref{eq:csl}, \eqref{eq:Gaml}, \eqref{eq:cst}, and~\eqref{eq:Gamt} with data given by the method of the Helfand moments~\cite{MG23_primo}.  For the great majority of the values, the agreement is excellent. The few discrepancies are explained by the fact that the linear regression is made over three points only and that the third point might fall already outside the hydrodynamic regime.

\subsection{Dispersion relations} 
\label{sec:DR}

The dispersion relations for the three directions   $[100]$, $[110]$, and $[111]$ are shown in figure~\ref{Fig:DR_1d037}  for density $n_*=1.037$ and  in figure~\ref{Fig:DR_1d3}  for density $n_*=1.3$. The symbols correspond to the values of the  $q$-dependent coefficients defined by equation~\eqref{eq:q-coefficients} and given in tables~\ref{Tab:CC1.037_100}-\ref{Tab:CC1.3_111}.

 The lines  depict the dispersion relations obtained from the asymptotic values of the   coefficients at $q=0$. These asymptotic values have been  evaluated with the linear regression on the $q^2$-dependence of the  coefficients~\eqref{eq:q-coefficients} at non-vanishing values of $q$.  As expected, the  coefficients can be approximated by a constant asymptotic value   when~$q_*$ is smaller than one, which is the value approximately delimiting the hydrodynamic regime. At larger values of~$q_*$, the dependence on the wave number cannot be neglected, which can be seen from the deviations between the actual values, shown with symbols in the figures, and the dispersion relations obtained with the asymptotic values of the coefficients extrapolated to $q=0$.  We note that, beyond the hydrodynamic regime, the real parts of the dispersion relations have a $q$-dependence that is smaller than linear, in agreement with the experimental observations of the dispersion relations by neutron inelastic scattering in crystals~\cite{SBR67,SROR72,TES78}.
 
 Moreover, figures~\ref{Fig:DR_1d037} and~\ref{Fig:DR_1d3} confirm the predictions shown in figures~5 and~6 of reference~\cite{MG23_secundo} for the dispersion relations in the hydrodynamic regime.  In particular, the real and imaginary parts of the dispersion relations are consistently larger for the longitudinal than the transverse sound waves, because $c_{\rm l} > c_{{\rm t}_{1,2}}$ and $\Gamma_{\rm l} > \Gamma_{{\rm t}_{1,2}}$, while $\Gamma_{\rm l}\sim\chi$.  Figures~\ref{Fig:DR_1d037} and~\ref{Fig:DR_1d3} also confirm the equality of the dispersion relations of the two transverse sound waves in the directions $[100]$ and $[111]$, and their difference in the direction $[110]$.
 
 In addition, the values of the quantities given in tables~\ref{Tab:CC1.037_100}-\ref{Tab:CC1.037_111} for the density $n_*=1.037$ and in tables~\ref{Tab:CC1.3_100}-\ref{Tab:CC1.3_111} for $n_*=1.3$ also show that the speeds and the diffusivities increase with the density.  These results are consistent with the behavior of these quantities, diverging as $c_\sigma,\Gamma_\sigma,\chi \sim (\sqrt{2}-n_*)^{-1}$ near the close-packing density $n_*=\sqrt{2}$ \cite{MG23_secundo}.

\section{From the poles to the elastic and transport coefficients}
\label{sec:ECTC}

In this section, we  evaluate the elastic and transport coefficients using the poles of the spectral functions and compare the results to the same coefficients obtained with the method of  Helfand moments~\cite{MG23_primo}. We assume that the equation of state for the hydrostatic pressure is known. The latter has been computed for a system of hard spheres in the solid phase in references~\cite{S98,MG23_primo}. From the equation of state, we obtain the specific heat ratio $\gamma$ and the isothermal bulk modulus $B_T$. The specific heat capacity at constant volume takes the value $c_v\equiv(\partial e_0/\partial T)_v=3k_{\rm B}/(2m)$. 

By analyzing the set of spectral functions in the three directions of table~\ref{Tab:LTCoeffs} for the wave vector, an over-constrained system of equations is obtained for the elastic and transport coefficients. To be more general, we restrict to the identification of the solutions that minimize the system without making any assumptions about the underlying data.

\subsection{Elastic coefficients}
\label{subec:EC}

From the extrapolated values of the poles at $q=0$, we first compute the longitudinal and transverse stress-strain coefficients $B^T_{\rm l}$, $B^T_{{\rm t}_1}$, and $B^T_{{\rm t}_2}$ in the three considered  directions using equations~\eqref{eq:csl} and~\eqref{eq:cst}. The results are given in tables~\ref{tab:LTC1d037} and~\ref{tab:LTC1d3} for the densities $n_*=1.037$ and $n_*=1.3$.  A comparison of the results with the values obtained by the method of  Helfand moments is given in the same tables, showing good agreement.

Next, using the definitions of the longitudinal and transverse isothermal stress-strain coefficients in table~\ref{Tab:LTCoeffs}, and the following relationship between the isothermal bulk modulus and  stress-strain coefficients
\begin{align}
B_T &=\frac{1}{3}\left(B^T_{11}+2B^T_{12}\right) ,
\label{eq:Cbulkm}
\end{align}
we obtain an over-constrained system of equations $\boldsymbol{\mathsf A}\cdot{\bf X} = {\bf B}$,
where 
\begin{align}
& {\bf X} \equiv
\left(
\begin{array}{l}
B^T_{11}  \\
B^T_{12}  \\
B^T_{44} \\
\end{array}
\right) , 
\qquad
\boldsymbol{\mathsf A} \equiv \left(
\begin{array}{c c c }
1/3  & 2/3 & 0  \\
1  & 0 & 0 \\
0  & 0 & 1  \\
1/2  & 1/2 & 1 \\
1/2  & -1/2 & 0  \\
0  & 0 & 1 \\
1/3  & 2/3 & 4/3  \\
1/3  & -1/3 & 1/3  \\
\end{array}
\right) ,
\qquad
{\bf B}\equiv \left(
\begin{array}{l} B_T \\ 
B^T_{{\rm l}}\vert_{[100]}\\ 
 B^T_{{\rm t}_{1,2}}\vert_{[100]}\\ 
  B^T_{{\rm l}}\vert_{[110]}\\ 
   B^T_{{\rm t}_{1}}\vert_{[110]}\\ 
    B^T_{{\rm t}_{2}}\vert_{[110]}\\ 
     B^T_{{\rm l}}\vert_{[111]} \\  
     B^T_{{\rm t}_{1,2}}\vert_{[111]}
\end{array}
\right) .
\end{align}
The  best approximate solution of this system, such that $\Vert\boldsymbol{\mathsf A}\cdot{\bf X}-{\bf B}\Vert$ is minimum, is given by the well-known formula ${\bf X} = (\boldsymbol{\mathsf A}^{\rm T}\cdot\boldsymbol{\mathsf A})^{-1}\cdot\boldsymbol{\mathsf A}^{\rm T}\cdot{\bf B}$~\cite{PT56}, which here gives the isothermal stress-strain coefficients. The isothermal elastic constants are finally obtained with the relations $C^T_{11}=B^T_{11}+p$, $C^T_{12}=B^T_{12}-p$, and $C^T_{44}=B^T_{44}+p$.

The results for the isothermal elastic constants are given in tables~\ref{tab:ECTC1d037} and~\ref{tab:ECTC1d3} for the densities $n_*=1.037$ and $n_*=1.3$. The results are in agreement with the elastic coefficients obtained with the method of Helfand moments. 
 
\subsection{Transport coefficients}
\label{subec:TC}
 
\paragraph{Viscosities.} The procedure for extracting the viscosities  $\eta_{11}$, $\eta_{12}$, and $\eta_{44}$  from the extrapolated pole values at $q=0$ is similar to that employed for determining the isothermal elastic constants. We first obtain the longitudinal and transverse viscosities $\eta_{\rm l}$, $\eta_{{\rm t}_1}$, and $\eta_{{\rm t}_2}$  in the three considered  directions using equations~\eqref{eq:Gaml},~\eqref{eq:chi}, and~\eqref{eq:Gamt}. For the densities $n_*=1.037$ and $n_*=1.3$, the results are respectively given in tables~\ref{tab:LTC1d037} and~\ref{tab:LTC1d3}, where they are also compared to the values obtained by the method of  Helfand moments.

Next, using the definitions of the longitudinal and transverse viscosities in table~\ref{Tab:LTCoeffs}, we obtain the  over-constrained system of equations $\boldsymbol{\mathsf A}^\prime\cdot{\bf Y} = {\bf H}$, where 
\begin{align}
& {\bf Y} \equiv
\left(
\begin{array}{l}
\eta_{11}  \\
\eta_{12}  \\
\eta_{44} \\
\end{array}
\right) , 
\qquad
\boldsymbol{\mathsf A}^\prime \equiv
\left(
\begin{array}{c c c }
1  & 0 & 0 \\
0  & 0 & 1  \\
1/2  & 1/2 & 1 \\
1/2  & -1/2 & 0  \\
0  & 0 & 1 \\
1/3  & 2/3 & 4/3  \\
1/3  & -1/3 & 1/3  \\
\end{array}
\right) ,
\qquad
{\bf H}\equiv \left(
\begin{array}{l} 
\eta_{{\rm l}}\vert_{[100]}\\ 
 \eta_{{\rm t}_{1,2}}\vert_{[100]}\\ 
  \eta_{{\rm l}}\vert_{[110]}\\ 
   \eta_{{\rm t}_{1}}\vert_{[110]}\\ 
    \eta_{{\rm t}_{2}}\vert_{[110]}\\ 
     \eta_{{\rm l}}\vert_{[111]} \\  
     \eta_{{\rm t}_{1,2}}\vert_{[111]}\end{array}
\right) .
\end{align}
As for the elastic constants, we look for  the best approximate solution of this system. The results for the isothermal viscosities are given in tables~\ref{tab:ECTC1d037} and~\ref{tab:ECTC1d3} for the densities $n_*=1.037$ and $n_*=1.3$. These results are in agreement with the viscosities obtained with the method of Helfand moments. 
 
 \paragraph{Heat conductivity.} In each direction, the heat conductivity is obtained from the thermal diffusivity $D_T$, which is extracted from the coefficient $\chi$  given by equation~\eqref{eq:chi}. We obtain three values  for the heat conductivity $\{\kappa^{(i)}\pm\Delta \kappa_{\rm P}^{(i)}\}$, where $i\in\{[100],[110],[111]\}$ and $\Delta \kappa_{\rm P}^{(i)}$ is the error  on $\kappa^{(i)}$, which comes from the uncertainty on the location of the poles. The heat conductivity is obtained as $\kappa=\sum_{i}\kappa^{(i)}/3$ and the  error on $\kappa$ is estimated as $\Delta \kappa=\sqrt{\sum_{i}(\Delta \kappa^{(i)})^2/6}$, where $\Delta \kappa^{(i)}=|\kappa^{(i)}-\kappa|+|\Delta \kappa^{(i)}_{\text{P}}|$. The results for the heat conductivity are given in tables~\ref{tab:ECTC1d037} and~\ref{tab:ECTC1d3} for the densities $n_*=1.037$ and $n_*=1.3$. The results are   also in agreement with the elastic coefficients obtained with the method of Helfand moments. 

 \section{Conclusion and perspectives}
 \label{sec:concl}
  
In this work, we have obtained the elastic and transport coefficients for a perfect crystal of hard spheres, from the poles of the hydrodynamic spectral functions. We have shown that the results are in agreement with the values obtained by the method of Helfand moments~\cite{MG23_primo}. Currently, the latter is more accurate and is computationally more efficient, since shorter time intervals can be considered. The sometimes large uncertainty on the elastic and transport coefficients when computed from the poles stems from the uncertainty on the location of the resonance peaks and on the extrapolation to $q=0$ with the least square linear regression. For the latter, only three points could be considered since we need to stay in the hydrodynamic regime.

In order to predict the transport coefficients from the poles with better accuracy, it would be necessary to reduce the magnitude of the wave number $q$, which would require looking at higher numbers of particles. By doing so, we would have more data points in the hydrodynamic regime for the poles, over which the extrapolation to finite $q$ is done. We would also expect a more precise estimation of the location of the poles since the peaks are typically  sharper when $q$ is smaller. However, as explained in section~\ref{subsec:MDHS}, increasing the number $N$ of particles also increases the number of collisions and, thus, the computational cost, in particular, since the time interval needed for the simulation is constrained by the damping rates of the correlation functions. To reduce the computational cost at large $N$, a possibility  would be to improve the event-driven algorithm, as in reference~\cite{S22}.

Given these limitations of the direct computational method to evaluate the elastic and transport coefficients from the poles of the spectral functions, the agreement demonstrated in tables~\ref{tab:ECTC1d037} and~\ref{tab:ECTC1d3} with the values obtained by the method of Helfand moments~\cite{MG23_primo} provides a validation of the local-equilibrium approach to the hydrodynamics of perfect crystals~\cite{MG23_secundo} in the case of the hard-sphere system.

Furthermore, tables~\ref{tab:ECTC1d037} and~\ref{tab:ECTC1d3} for the densities $n_*=1.037$ and $n_*=1.3$ also show that the elastic and transport coefficients increase with the particle density in consistency with the results obtained in reference~\cite{MG23_primo}, according to which the elastic coefficients diverge as $C^T_\mu\sim (\sqrt{2}-n_*)^{-2}$ and the transport coefficients as $\kappa,\eta_\mu\sim(\sqrt{2}-n_*)^{-1}$ (with $\mu\in\{11,12,44\}$) near the close-packing density $n_*=\sqrt{2}$ in the hard-sphere crystal.

In addition, the results indicate that the viscosities of crystals as well as their heat conductivities can be estimated by measuring the widths of the resonances, for example, in neutron inelastic scattering spectra commonly used to determine the dispersion relations of sound waves in solids~\cite{SBR67,SROR72,TES78}.  

The present study of hydrodynamics in perfect crystals~\cite{MG23_primo,MG23_secundo} leaves open the issue of the eighth hydrodynamic mode of vacancy diffusion.  In general, this extra mode is also contributing to the irreversible transport processes generating dissipation in crystals.  The study of this mode is challenging because of its intricacies with the periodicity and elasticity of the crystal.  We hope to address this issue in future work.


\section*{Acknowledgements}

The authors acknowledge the support of the Universit\'e Libre de Bruxelles (ULB) and the Fonds de la Recherche Scientifique de Belgique (F.R.S.-FNRS) in this research. J.~M. is a Postdoctoral Researcher of the Fonds de la Recherche Scientifique de Belgique (F.R.S.-FNRS).  Computational resources have been provided by the Consortium des Equipements de Calcul Intensif (CECI), funded by the Fonds de la Recherche Scientifique de Belgique (F.R.S.-FNRS) under Grant No. 2.5020.11 and by the Walloon Region.

 \appendix
 

\section{Extrapolation to $q=0$}
\label{app:extq0}

To  extrapolate the values of the coefficients~\eqref{eq:q-coefficients} to $q=0$, a linear least square fitting is performed on the data as function of $q^{2}$. We consider data points $\{x_i,y_i\}_{i=1}^M$ with $x_i=q^2_i$, as shown in figures~\ref{Fig:LR_rho1.037} and~\ref{Fig:LR_rho1.3}. The fluctuations of the values $y_i$ are denoted by $\Delta y_i$, but there are no fluctuations on the values $x_i$.  Linear least square fitting is obtained from the minimum of  $R = \frac{1}{2} \sum_{i=1}^M (y_i-a-b x_i)^2$. Accordingly, the line $y=a+bx$ that is fitted to the data has the slope $b$ and the ordinate at origin $a$, which are given by
\begin{align}
& b = \frac{\overline{x\, y}-\bar x\, \bar y}{\overline{x^2} - {\bar x}^2} 
\qquad\mbox{and}\qquad
a = \bar y -  \bar x \, b \, , \qquad
\mbox{where} \qquad 
\overline{(\cdot)} \equiv \frac{1}{M} \, \sum_{i=1}^M (\cdot)\,.
\end{align}
 The error on the ordinate at origin can be estimated with
 \begin{align}
\Delta a \simeq \sqrt{\frac{1}{M} \sum_{i=1}^M \left( \frac{ {\bar x}^2-\bar x \, x_i}{\overline{x^2} - {\bar x}^2}\right)^2 \Delta y_i ^2 \ }.
\end{align}


\pagebreak


\begin{table}[h!]
  \begin{tabular}{c @{\hskip 1cm} c @{\hskip 1cm} c @{\hskip 1cm} c @{\hskip 1cm} c @{\hskip 1cm} c }
    \hline\hline
    $q_*$  &     $[\chi(q)]_*$  & $[c_{\rm l}(q)]_*$   &      $[\Gamma_{\rm l}(q)]_*$      &  $[c_{{\rm t}_{1,2}}(q)]_*$ &  $[\Gamma_{{\rm t}_{1,2}}(q)]_*$	\\
    \hline   
    1.50&$ 2.53\pm0.11$& $11.21 \pm0.13 $&$ 3.49\pm0.10$&$ 5.44\pm0.02 $&$ 1.88\pm0.03$ \\
    1.00&$ 2.97\pm0.05 $ & $11.56 \pm0.07 $&$ 3.78\pm0.07 $&$ 5.58\pm0.03 $&$ 1.85\pm0.01$ \\
    0.50&$ 3.34\pm0.12 $ & $11.81 \pm0.01 $&$ 4.02\pm0.02 $&$ 5.62\pm0.01 $&$ 1.96\pm0.01$ \vspace{0.1cm}\\
    \vspace{0.1cm}
    0 (Lin. Reg.) &$ 3.41 \pm 0.19 $ & $11.88\pm0.09 $&$ 4.07\pm0.08 $&$ 5.65\pm0.02 $&$ 1.93\pm0.02$\\
    0 (Helfand) &$ 3.56\pm0.16$ & $11.86\pm0.04 $&$ 4.10\pm0.12 $&$ 5.64\pm0.18 $&$ 2.03\pm0.03$\\
    \hline\hline
      \end{tabular}
  \caption{Dependence on the wave number $q$ along the $[100]$ direction for the speeds of the longitudinal and transverse  sound waves and the diffusivities obtained from the poles of the spectral functions $S(q,\omega)$ and $J_{{\rm t}_{1,2}}(q,\omega)$ given by the numerical Fourier transforms of the intermediate scattering function and  the correlation functions of transverse momentum density fluctuations  at density $n_*=1.037$ for a solid of $N=2048$ hard spheres. The reported error is on the estimation of the location of the poles.  The penultimate row (Lin. Reg.) corresponds to the limit $q\rightarrow 0$ and is obtained using a linear least square  regression with equal weights on the data points for the different values of $q$. The reported error is the standard error on the fitted parameter. The last row (Helfand) is the corresponding value obtained from the Helfand moments~\cite{MG23_primo,MG23_secundo}.}\label{Tab:CC1.037_100}
\end{table}


\begin{table}[h!]
  \begin{tabular}{c @{\hskip 0.5cm} c @{\hskip 0.5cm} c @{\hskip 0.5cm} c @{\hskip 0.5cm} c @{\hskip 0.5cm} c @{\hskip 0.5cm} c @{\hskip 0.5cm} c }
    \hline\hline
    $q_*$  &     $[\chi(q)]_*$ & $[c_{\rm l}(q)]_*$   &      $[\Gamma_{\rm l}(q)]_*$    &  $[c_{{\rm t}_1}(q)]_*$ &  $[\Gamma_{{\rm t}_1}(q)]_*$&  $[c_{{\rm t}_2}(q)]_*$ &  $[\Gamma_{{\rm t}_2}(q)]_*$	\\
    \hline   
    2.12&$ 2.78\pm0.09  $& $9.82 \pm0.20 $&$ 2.87\pm0.07$ & $3.43\pm0.09$ & $0.84\pm0.02$ & $5.16\pm0.01$ & $1.62\pm0.01$\\
    1.42 &$ 3.71\pm0.11 $& $11.19 \pm0.19 $&$ 4.06\pm0.04 $ & $3.55\pm0.01$ & $0.96\pm0.01$ & $5.42\pm0.01$ & $1.80\pm0.01$\\
    0.71 &$ 4.06\pm0.24$& $12.23 \pm0.03 $&$ 4.66\pm0.05 $& $3.72\pm0.01$ & $1.01\pm0.01$ & $5.64\pm0.04$ & $2.01\pm0.02$\\
    \vspace{0.1cm}
    0 (Lin. Reg.) &$ 4.28 \pm 0.37$& $12.47\pm0.18 $&$ 4.91\pm0.09 $& $3.72\pm0.05$ & $1.04\pm0.01$ & $5.68\pm0.06$ & $2.03\pm0.03$ \\
    0 (Helfand) &$ 4.13\pm0.19$ & $12.59\pm0.08$&$  4.80\pm0.12$ & $3.73\pm0.06$ & $1.04\pm0.04$ & $5.64\pm0.18$ & $2.03\pm0.03$ \\
    \hline\hline
      \end{tabular}
  \caption{Dependence on the wave number $q$ along the $[110]$ direction for the speeds of the longitudinal and transverse sound waves and  the diffusivities obtained from the poles of the spectral functions $S(q,\omega)$ and $J_{{\rm t}_{1,2}}(q,\omega)$ given by the numerical Fourier transforms of the intermediate scattering function and  the correlation functions of transverse momentum density fluctuations  at density $n_*=1.037$ for a solid of $N=2048$ hard spheres. The reported error is on the estimation of the location of the poles.  The penultimate row (Lin. Reg.) corresponds to the limit $q\rightarrow 0$ and is obtained using a linear least square  regression with equal weights on the data points for the different values of $q$. The reported error is the standard error on the fitted parameter. The last row (Helfand) is the corresponding value obtained from the Helfand moments~\cite{MG23_primo,MG23_secundo}.}\label{Tab:CC1.037_110}
\end{table}
\begin{table}[h!]
  \begin{tabular}{c @{\hskip 1cm} c @{\hskip 1cm} c @{\hskip 1cm} c @{\hskip 1cm} c @{\hskip 1cm} c }
    \hline\hline
    $q_*$  &     $[\chi(q)]_*$    	& $[c_{\rm l}(q)]_*$   &      $[\Gamma_{\rm l}(q)]_*$      &  $[c_{{\rm t}_{1,2}}(q)]_*$ &  $[\Gamma_{{\rm t}_{1,2}}(q)]_*$\\
    \hline   
    2.60 & $ 2.72\pm0.01$& $8.72 \pm0.12 $&$ 3.18\pm0.07$ &$3.34\pm0.02$& $0.78\pm0.01$ \\
    1.73 & $ 3.13\pm0.21$& $10.68 \pm0.34 $&$ 4.41\pm0.06 $ & $3.96\pm0.01$ & $1.09\pm0.01$ \\
    0.87 & $ 4.15\pm0.13 $& $12.64 \pm0.07 $&$ 4.87\pm0.09 $ & $4.33\pm0.01$ & $1.30\pm0.01$\\
    \vspace{0.1cm}
    0 (Lin. Reg.) & $ 4.12 \pm 0.25$& $12.91\pm0.28 $&$ 5.15\pm0.15 $ & $4.45\pm0.01$ & $1.36\pm0.01$\\
    0 (Helfand) & $ 4.30\pm0.21$& $12.83\pm0.11$&$ 5.04\pm0.12 $ & $4.46\pm0.09$ & $1.37\pm0.03$ \\
    \hline\hline
      \end{tabular}
  \caption{Dependence on the wave number $q$ along the $[111]$ direction for the speeds of the longitudinal and transverse  sound waves and  the diffusivities  obtained from the poles of the spectral functions $S(q,\omega)$ and $J_{{\rm t}_{1,2}}(q,\omega)$ given by the numerical Fourier transforms of the intermediate scattering function and  the correlation functions of transverse momentum density fluctuations  at density $n_*=1.037$ for a solid of $N=2048$ hard spheres. The reported error is on the estimation of the location of the poles.  The penultimate row (Lin. Reg.) corresponds to the limit $q\rightarrow 0$ and is obtained using a linear least square  regression with equal weights on the data points for the different values of $q$. The reported error is the standard error on the fitted parameter. The last row (Helfand) is the corresponding value obtained from the Helfand moments~\cite{MG23_primo,MG23_secundo}.}\label{Tab:CC1.037_111}
\end{table}

\begin{table}[h!]
  \begin{tabular}{c @{\hskip 1cm} c @{\hskip 1cm} c @{\hskip 1cm} c @{\hskip 1cm} c  @{\hskip 1cm} c}
    \hline\hline
    $q_*$  &     $[\chi(q)]_*$    & $[c_{\rm l}(q)]_*$   &      $[\Gamma_{\rm l}(q)]_*$      &  $[c_{{\rm t}_{1,2}}(q)]_*$&  $[\Gamma_{{\rm t}_{1,2}}(q)]_*$	\\
    \hline   
    1.62 & $ 9.48\pm0.34$ & $39.51 \pm0.48 $&$ 8.89\pm0.30$&$ 22.25\pm0.75 $&$ 5.02\pm0.12$ \\
    1.08 & $ 12.33\pm0.09 $ & $41.36 \pm0.06 $&$ 9.93\pm0.05 $&$ 22.93\pm0.42 $&$ 6.39\pm0.31$ \\
    0.54 & $ 12.78\pm0.65 $ & $41.70 \pm0.04 $& $10.77\pm0.20 $&$ 23.16\pm0.05$&$ 6.84\pm0.04 $\\
    \vspace{0.1cm}
    0 (Lin. Reg.) & $ 13.53 \pm0.98 $ & $42.18\pm0.25$ &$ 10.95\pm0.34$ &$ 23.32\pm0.49$ &$ 7.17\pm0.25$\\
    0 (Helfand) & $ 13.49\pm0.39$ & $41.97\pm0.06$& $11.24\pm 0.22$&$23.09\pm0.12 $&$6.91\pm0.20$\\
    \hline\hline
      \end{tabular}
  \caption{Dependence on the wave number $q$ along the $[100]$ direction for the speeds of the longitudinal and transverse sound waves and  the diffusivities obtained from the poles of the spectral functions $S(q,\omega)$ and $J_{{\rm t}_{1,2}}(q,\omega)$ given by the numerical Fourier transforms of the intermediate scattering function and  the correlation functions of transverse momentum density fluctuations  at density $n_*=1.3$ for a solid of $N=2048$ hard spheres. The reported error is on the estimation of the location of the poles.  The penultimate row (Lin. Reg.) corresponds to the limit $q\rightarrow 0$ and is obtained using a linear least square  regression with equal weights on the data points for the different values of $q$. The reported error is the standard error on the fitted parameter. The last row (Helfand) is the corresponding value obtained from the Helfand moments~\cite{MG23_primo,MG23_secundo}.}\label{Tab:CC1.3_100}
\end{table}


\begin{table}[h!]
  \begin{tabular}{c @{\hskip 0.5cm} c @{\hskip 0.5cm} c @{\hskip 0.5cm} c @{\hskip 0.5cm} c @{\hskip 0.5cm} c @{\hskip 0.5cm} c @{\hskip 0.5cm} c }
    \hline\hline
    $q_*$  &     $[\chi(q)]_*$    & $[c_{\rm l}(q)]_*$   &      $[\Gamma_{\rm l}(q)]_*$      &  $[c_{{\rm t}_1}(q)]_*$ &  $[\Gamma_{{\rm t}_1}(q)]_*$&  $[c_{{\rm t}_2}(q)]_*$ &  $[\Gamma_{{\rm t}_2}(q)]_*$ 	\\
    \hline     
    2.29 &$ 5.91\pm0.06$&$ 33.54 \pm1.04 $&$ 9.07\pm0.62$ & $16.43\pm0.01$ &$2.43\pm0.01$ &$22.47\pm0.12$ & $5.35\pm0.04$\\
    1.53 &$ 10.39\pm0.04$& $39.55 \pm0.11 $&$ 11.55\pm0.03 $ &$17.00\pm0.05$ & $2.83\pm0.04$& $22.50\pm0.54$& $5.87\pm0.16$\\
     0.76 &$ 14.56\pm0.19$&$ 44.28\pm0.05 $&$ 12.99\pm0.06 $ &$17.33\pm0.03$ &$3.08\pm0.05$ & $23.14\pm0.02$&$6.90\pm0.04$\\
    \vspace{0.1cm}
    0 (Lin. Reg.) &$ 15.25 \pm 0.29$ & $45.33\pm0.52 $&$ 13.49\pm0.32$ & $17.45\pm0.06$&$3.16\pm0.08$ &$23.06\pm0.40$ &$6.90\pm 0.14$ \\
    0 (Helfand) &$ 15.13\pm 0.44$ & $44.65\pm0.07$&$ 14.19 \pm0.28$&$17.37\pm0.08$ &$3.14\pm0.05$ &$23.09\pm0.12$ &$6.91\pm0.20$\\
    \hline\hline
      \end{tabular}
  \caption{Dependence on the wave number $q$ along the $[110]$ direction for the speeds of the longitudinal and transverse sound waves and the diffusivities  obtained from the poles of the spectral functions $S(q,\omega)$ and $J_{{\rm t}_{1,2}}(q,\omega)$ given by the numerical Fourier transforms of the intermediate scattering function and  the correlation functions of transverse momentum density fluctuations  at density $n_*=1.3$ for a solid of $N=2048$ hard spheres. The reported error is on the estimation of the location of the poles.  The penultimate row (Lin. Reg.) corresponds to the limit $q\rightarrow 0$ and is obtained using a linear least square  regression with equal weights on the data points for the different values of $q$. The reported error is the standard error on the fitted parameter. The last row (Helfand) is the corresponding value obtained from the Helfand moments~\cite{MG23_primo,MG23_secundo}. }\label{Tab:CC1.3_110}
\end{table}
\begin{table}[h!]
  \begin{tabular}{c @{\hskip 1cm} c @{\hskip 1cm} c @{\hskip 1cm} c @{\hskip 1cm} c @{\hskip 1cm} c }
    \hline\hline
    $q_*$  &     $[\chi(q)]_*$    & $[c_{\rm l}(q)]_*$   &      $[\Gamma_{\rm l}(q)]_*$   	&  $[c_{{\rm t}_{1,2}}(q)]_*$&  $[\Gamma_{{\rm t}_{1,2}}(q)]_*$\\
    \hline   
    2.81 &$ 6.16\pm0.88$& $26.71 \pm 0.84 $&$ 10.62\pm0.26$ &$15.26\pm0.05$&$2.27\pm0.02$ \\
    1.87 &$ 10.53\pm0.77$ & $36.58 \pm 1.08 $&$ 13.29\pm0.20$ &$17.89\pm0.04$&$3.11\pm0.01$\\
    0.94 &$ 14.81\pm0.36$ & $44.77 \pm 0.39 $&$ 14.58\pm0.39 $ &$19.39\pm0.02$&$3.97\pm0.02$\\
    \vspace{0.1cm}
    0 (Lin. Reg.) &$ 15.44 \pm 0.89$ & $46.42 \pm1.07$&$ 15.16\pm0.61 $&$19.93\pm0.05$&$4.09\pm0.04$ \\
    0 (Helfand) &$ 15.59\pm0.45 $& $45.50 \pm0.09$&$ 15.21\pm0.33 $&$19.46\pm0.07$&$4.39\pm0.08$\\
    \hline\hline
      \end{tabular}
  \caption{Dependence on the wave number $q$ along the $[111]$ direction for the speeds of the longitudinal and transverse sound waves and the diffusivities  obtained from the poles of the spectral functions $S(q,\omega)$ and $J_{{\rm t}_{1,2}}(q,\omega)$ given by the numerical Fourier transforms of the intermediate scattering function and  the correlation functions of transverse momentum density fluctuations  at density $n_*=1.3$ for a solid of $N=2048$ hard spheres. The reported error is on the estimation of the location of the poles.  The penultimate row (Lin. Reg.) corresponds to the limit $q\rightarrow 0$ and is obtained using a linear least square  regression with equal weights on the data points for the different values of $q$. The reported error is the standard error on the fitted parameter. The last row (Helfand) is the corresponding value obtained from the Helfand moments~\cite{MG23_primo,MG23_secundo}.}\label{Tab:CC1.3_111}
\end{table}

\begin{table}[h!]
  \begin{tabular}{l @{\hskip 1.cm} c @{\hskip 1,cm} c }
    \hline\hline
      	 & Poles   &      Helfand     	\\
    \hline   
   $\left.B^T_{{\rm l}*}\right|_{[100]}$ & $60.59 \pm 2.14  $&$  60.11\pm0.88   $   \\
   $\left.B^T_{{\rm t}_{1,2}*}\right|_{[100]}$ & $ 33.14 \pm  0.29 $  & $33.03 \pm  2.15$ \\
    $\left.B^T_{{\rm l}*}\right|_{[110]}$ & $75.58 \pm 4.54 $ & $ 78.70 \pm 2.20 $ \\
    $\left.B^T_{{\rm t}_{1}*}\right|_{[110]}$ & $14.38 \pm 0.36  $&$ 14.45 \pm 0.49 $  \\
    $\left.B^T_{{\rm t}_{2}*}\right|_{[110]}$ & $ 33.49 \pm 0.71 $  & $ 33.03 \pm 2.15 $\\
    $\left.B^T_{{\rm l}*}\right|_{[111]}$ & $87.25 \pm 7.41 $ & $ 84.89  \pm  2.89$\\
    $\left.B^T_{{\rm t}_{1,2}*}\right|_{[111]}$ & $20.57\pm 0.23$  & $20.64 \pm 0.79 $\\
     $\left.\eta_{{\rm l}*}\right|_{[100]}$ & $3.43 \pm 0.35  $&$ 3.24 \pm 0.10   $   \\
   $\left.\eta_{{\rm t}_{1,2}*}\right|_{[100]}$ & $ 4.01 \pm  0.05 $  & $4.21 \pm 0.06$ \\
    $\left.\eta_{{\rm l}*}\right|_{[110]}$ & $5.16 \pm 0.52 $ & $ 5.28 \pm 0.10 $ \\
    $\left.\eta_{{\rm t}_{1}*}\right|_{[110]}$ & $2.16 \pm 0.03  $&$  2.17 \pm  0.08$  \\
    $\left.\eta_{{\rm t}_{2}*}\right|_{[110]}$ & $ 4.21 \pm 0.06 $  & $ 4.21 \pm  0.06$\\
    $\left.\eta_{{\rm l}*}\right|_{[111]}$ & $ 6.49 \pm 0.47 $ & $  5.96\pm  0.12$\\
    $\left.\eta_{{\rm t}_{1,2}*}\right|_{[111]}$ & $2.83\pm 0.02$  & $ 2.85\pm 0.06$\\
    \hline\hline
      \end{tabular}
  \caption{Coefficients $B_{\rm l}^T$, $B_{{\rm t}_{1,2}}^T$, $\eta_{\rm l}$, $\eta_{{\rm t}_{1,2}}$ of table~\ref{Tab:LTCoeffs} for the directions $[100]$, $[110]$, and $[111]$  at density $n_*=1.037$  obtained from the poles of the spectral functions of a solid of $N=2048$ hard spheres, compared to the values obtained by the method of Helfand moments in reference~\cite{MG23_primo}.}\label{tab:LTC1d037}
\end{table}

\begin{table}[h!]
  \begin{tabular}{l @{\hskip 1.cm} c @{\hskip 1,cm} c }
    \hline\hline
      	 & Poles   &      Helfand     	\\
    \hline   
   $\left.B^T_{{\rm l}*}\right|_{[100]}$ & $1142\pm 27  $&$ 1120 \pm 6  $   \\
   $\left.B^T_{{\rm t}_{1,2}*}\right|_{[100]}$ & $ 707 \pm  30 $  & $ 693 \pm  7$ \\
    $\left.B^T_{{\rm l}*}\right|_{[110]}$ & $1500 \pm 62 $ & $  1420\pm 8 $ \\
    $\left.B^T_{{\rm t}_{1}*}\right|_{[110]}$ & $396 \pm 3  $&$ 392 \pm 4 $  \\
    $\left.B^T_{{\rm t}_{2}*}\right|_{[110]}$ & $ 691 \pm 24 $  & $  693 \pm 7 $\\
    $\left.B^T_{{\rm l}*}\right|_{[111]}$ & $1630 \pm 130 $ & $ 1521 \pm 10  $\\
    $\left.B^T_{{\rm t}_{1,2}*}\right|_{[111]}$ & $516\pm 2$  & $493 \pm 3$\\
     $\left.\eta_{{\rm l}*}\right|_{[100]}$ & $10.45 \pm 1.61  $&$10.90  \pm  0.19 $   \\
   $\left.\eta_{{\rm t}_{1,2}*}\right|_{[100]}$ & $ 18.65 \pm  0.64 $  & $ 17.95 \pm0.53  $ \\
    $\left.\eta_{{\rm l}*}\right|_{[110]}$ & $19.61 \pm 0.99 $ & $ 20.69 \pm 0.54 $ \\
    $\left.\eta_{{\rm t}_{1}*}\right|_{[110]}$ & $8.21 \pm 0.22$&$ 8.16 \pm 0.14 $  \\
    $\left.\eta_{{\rm t}_{2}*}\right|_{[110]}$ & $ 17.94 \pm 0.35 $  & $ 17.95 \pm 0.53 $\\
    $\left.\eta_{{\rm l}*}\right|_{[111]}$ & $ 25.00 \pm 1.97 $ & $ 23.95 \pm 0.72 $\\
    $\left.\eta_{{\rm t}_{1,2}*}\right|_{[111]}$ & $10.63\pm 0.09$  & $ 11.43\pm 0.20$\\
    \hline\hline
      \end{tabular}
  \caption{Coefficients $B_{\rm l}^T$, $B_{{\rm t}_{1,2}}^T$, $\eta_{\rm l}$, $\eta_{{\rm t}_{1,2}}$ of table~\ref{Tab:LTCoeffs} for the directions $[100]$, $[110]$, and $[111]$  at density $n_*=1.3$  obtained from the poles of the spectral functions of a solid of $N=2048$ hard spheres, compared to the values obtained by the method of Helfand moments in reference~\cite{MG23_primo}. }\label{tab:LTC1d3}
\end{table}

\begin{table}[h!]
  \begin{tabular}{c @{\hskip 1.cm} c @{\hskip 1,cm} c }
    \hline\hline
      	 & Poles   &      Helfand     	\\
    \hline   
   $C^T_{11*}$ & $70.61 \pm 1.73  $&$ 71.66 \pm 0.88  $   \\
   $C^T_{12*}$ & $19.96 \pm  1.37 $  & $19.67 \pm 0.44 $ \\
    $C^T_{44*}$ & $45.49 \pm 3.78 $ & $ 44.58 \pm 2.15 $ \\
    $\eta_{11*}$ & $3.32 \pm 0.31  $&$ 3.24 \pm 0.10 $  \\
    $\eta_{12*}$ & $ -0.82 \pm 0.28 $  & $ -1.10 \pm 0.12 $\\
    $\eta_{44*}$ & $4.32 \pm 0.21 $ & $ 4.21 \pm 0.06 $\\
    $\kappa_{*}$ & $13.23\pm0.59$  & $13.43 \pm 0.60$\\
    \hline\hline
      \end{tabular}
  \caption{Elastic and transport coefficients at density $n_*=1.037$ obtained from the poles of the spectral functions of a solid of $N=2048$ hard spheres, compared to the values obtained by the method of Helfand moments in reference~\cite{MG23_primo}. }\label{tab:ECTC1d037}
\end{table}
\begin{table}[h!]
  \begin{tabular}{c @{\hskip 1.cm} c @{\hskip 1,cm} c }
    \hline\hline
      	 & Poles   &      Helfand     	\\
    \hline   
   $C^T_{11*}$ & $ 1184 \pm 28  $&$ 1168\pm 6  $   \\
   $C^T_{12*}$ & $  296 \pm 24 $  & $287 \pm 3  $ \\
    $C^T_{44*}$ & $807 \pm 66 $ & $ 741 \pm 7 $ \\
    $\eta_{11*}$ & $9.99 \pm 0.97  $&$ 10.90 \pm 0.19 $  \\
    $\eta_{12*}$ & $-4.77 \pm 0.94 $  & $ -5.43 \pm 0.20  $\\
    $\eta_{44*}$ & $18.26 \pm 0.86 $ & $17.96 \pm 0.53 $\\
    $\kappa_{*}$ & $ 52.70\pm 1.81$  & $53.81 \pm 1.55$\\
    \hline\hline
      \end{tabular}
  \caption{Elastic and transport coefficients at density $n_*=1.037$ obtained from the poles of the spectral functions of a solid of $N=2048$ hard spheres, compared to the values obtained by the method of Helfand moments in reference~\cite{MG23_primo}. }\label{tab:ECTC1d3}
\end{table}


\begin{figure}[h!]\centering
{\includegraphics[width=1.\textwidth]{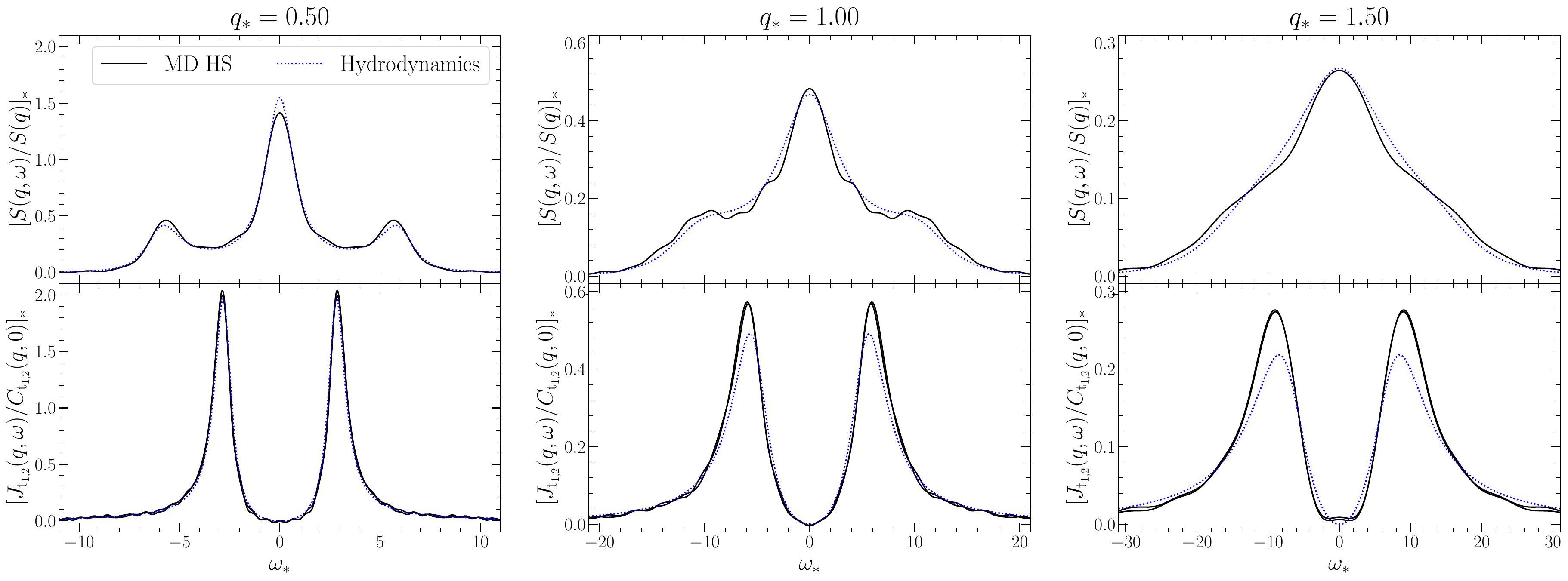}}
\caption[] {Normalized dynamic structure factor $S(q,\omega)$ and  spectral functions of transverse momentum density fluctuations $J_{{\rm t}_{1,2}}(q,\omega)$ versus the frequency at density $n_*=1.037$ for a solid of $N=2048$ hard spheres for ${\bf q}$ in the direction $[100]$ with magnitudes $q_*\in\{0.50, 1.00, 1.50\}$. The solid lines correspond to the spectral functions obtained from molecular dynamics simulations of hard spheres. The dotted lines correspond to the analytical expressions predicted by hydrodynamics in equations~\eqref{eq:DSF_RBP} and \eqref{eq:JtoCt}, with the parameters obtained in references~\cite{MG23_primo,MG23_secundo}.}\label{Fig:CF100-1.037}
\end{figure}


\begin{figure}[h!]\centering
{\includegraphics[width=1.\textwidth]{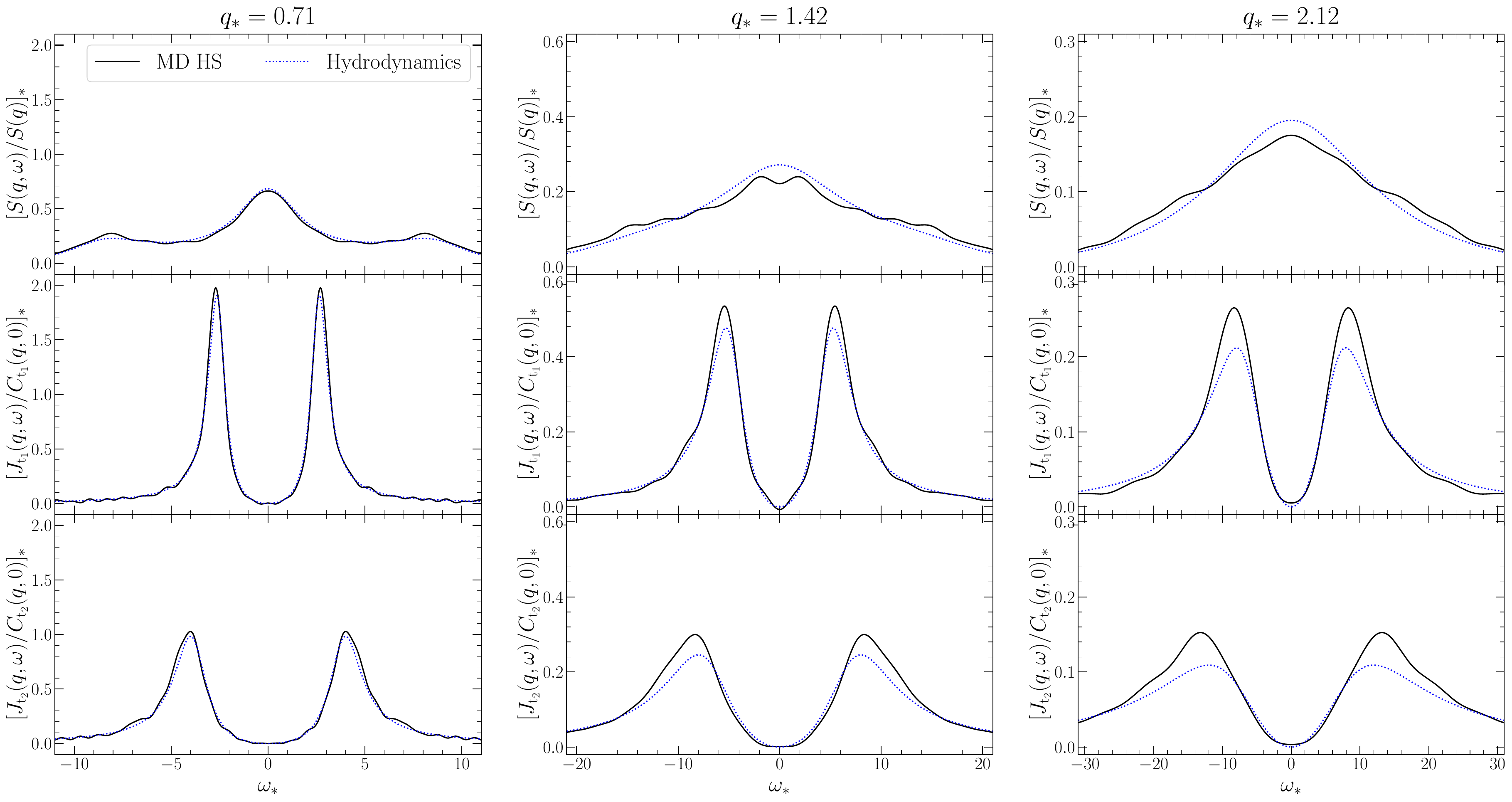}}
\caption[] {Normalized dynamic structure factor $S(q,\omega)$ and   spectral functions of transverse momentum density fluctuations $J_{{\rm t}_{1,2}}(q,\omega)$ versus the frequency at density $n_*=1.037$ for a solid of $N=2048$ hard spheres for ${\bf q}$ in the direction $[110]$ with magnitudes $q_*\in\{0.71, 1.41, 2.12\}$. The solid lines correspond to the spectral functions obtained from molecular dynamics simulations of hard spheres. The dotted lines correspond to the analytical expressions predicted by hydrodynamics in equations~\eqref{eq:DSF_RBP} and \eqref{eq:JtoCt}, with the parameters obtained in references~\cite{MG23_primo,MG23_secundo}.}\label{Fig:CF110-1.037}
\end{figure}


\begin{figure}[h!]\centering
{\includegraphics[width=1.\textwidth]{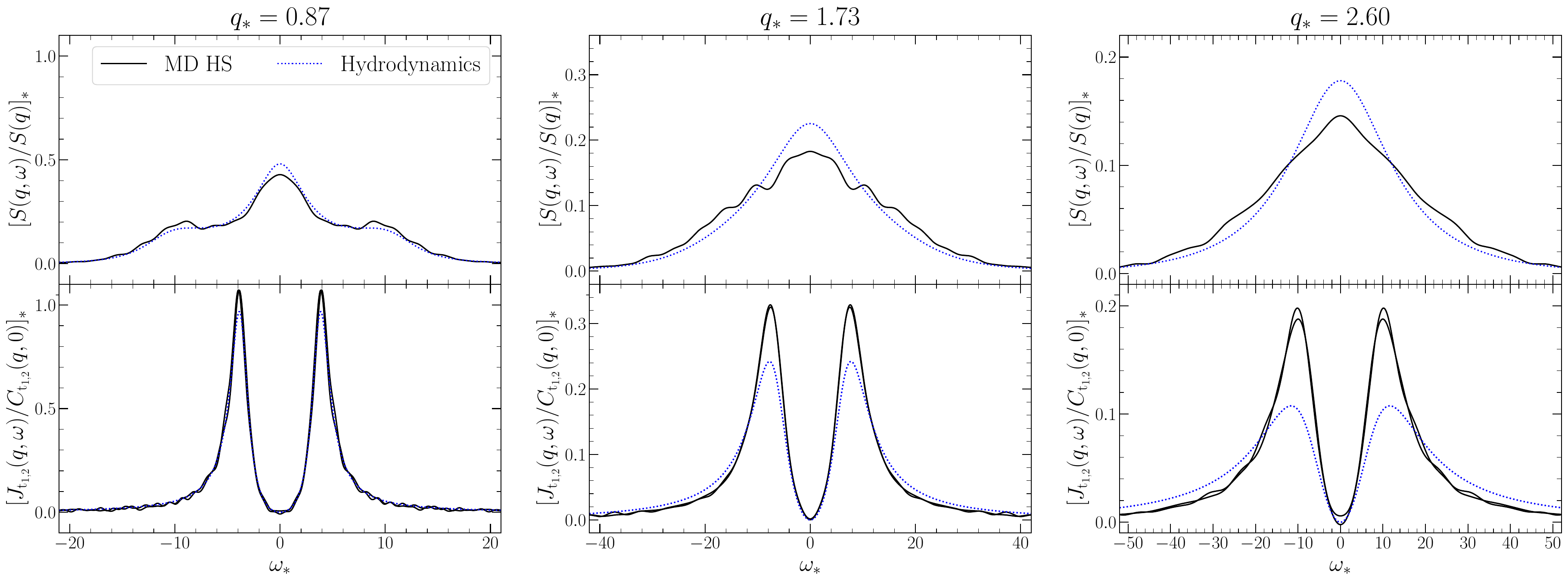}}
\caption[] {Normalized dynamic structure factor $S(q,\omega)$ and   spectral functions of   transverse momentum density fluctuations $J_{{\rm t}_{1,2}}(q,\omega)$ versus the frequency at density $n_*=1.037$ for a solid of $N=2048$ hard spheres for ${\bf q}$ in the direction $[111]$ with magnitudes $q_*\in\{0.87, 1.73, 2.60\}$. The solid lines correspond to the spectral functions obtained from  molecular dynamics simulations of hard spheres. The dotted lines correspond to the analytical expressions predicted by hydrodynamics in equations~\eqref{eq:DSF_RBP} and \eqref{eq:JtoCt}, with the parameters obtained in references~\cite{MG23_primo,MG23_secundo}.}\label{Fig:CF111-1.037}
\end{figure}


\begin{figure}[h!]\centering
{\includegraphics[width=1.\textwidth]{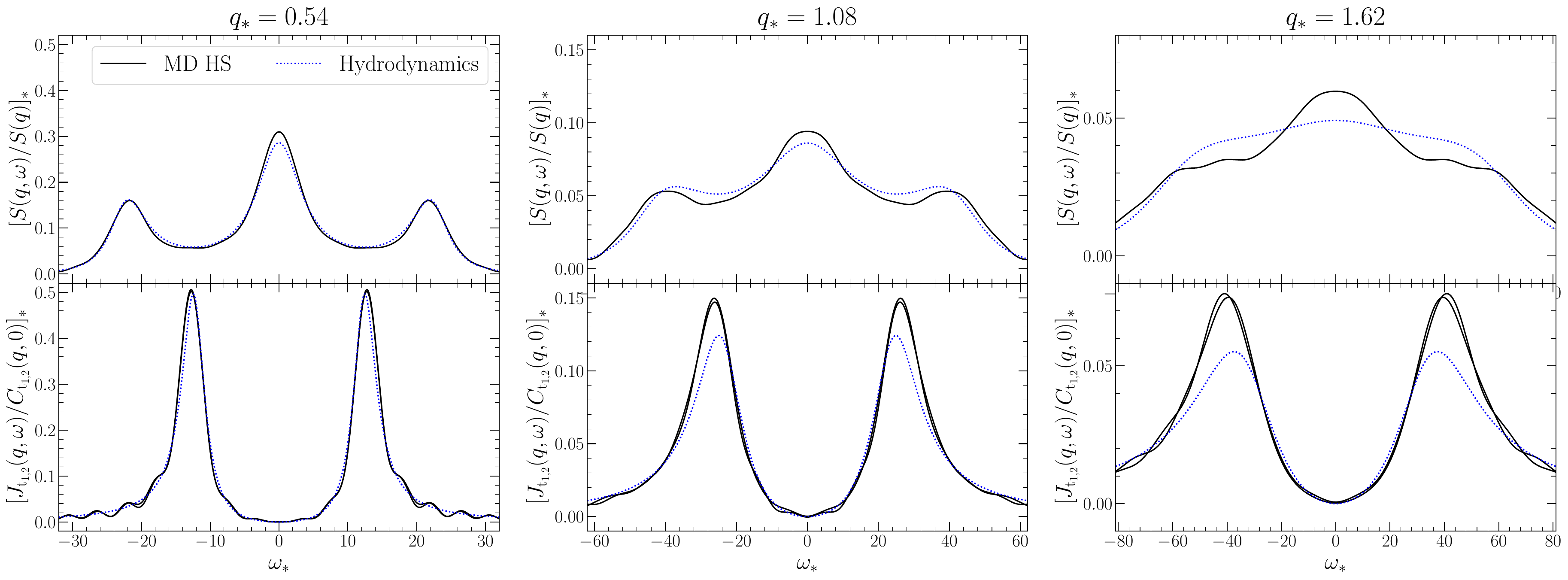}}
\caption[] {Normalized dynamic structure factor $S(q,\omega)$ and  spectral functions of  transverse momentum density fluctuations $J_{{\rm t}_{1,2}}(q,\omega)$ versus the frequency at density $n_*=1.3$ for a solid of $N=2048$ hard spheres for ${\bf q}$ in the direction $[100]$ with magnitudes $q_*\in\{0.54, 1.08, 1.62\}$. The solid lines correspond to the spectral functions obtained from molecular dynamics simulations of hard spheres. The dotted lines correspond to the analytical expressions predicted by hydrodynamics in equations~\eqref{eq:DSF_RBP} and \eqref{eq:JtoCt}, with the parameters obtained in references~\cite{MG23_primo,MG23_secundo}.}\label{Fig:CF100-1.3}
\end{figure}


\begin{figure}[h!]\centering
{\includegraphics[width=1.\textwidth]{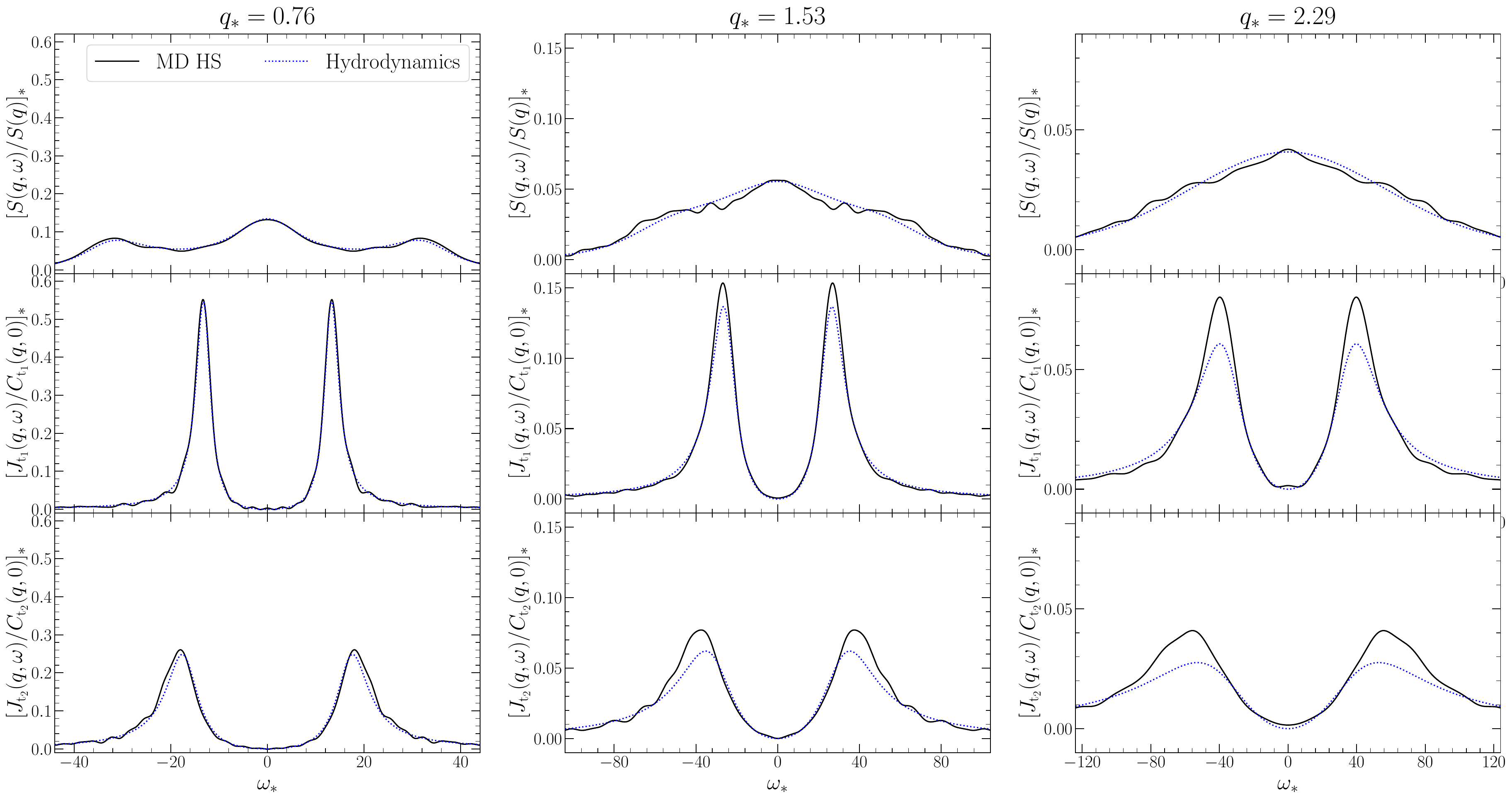}}
\caption[] {Normalized dynamic structure factor $S(q,\omega)$ and  spectral functions of  transverse momentum density fluctuations $J_{{\rm t}_{1,2}}(q,\omega)$ versus the frequency at density $n_*=1.3$ for a solid of $N=2048$ hard spheres for ${\bf q}$ in the direction $[110]$ with magnitudes $q_*\in\{0.76, 1.53, 2.29\}$. The solid lines correspond to the spectral functions obtained from   molecular dynamics simulations of hard spheres. The dotted lines correspond to the analytical expressions predicted by hydrodynamics in equations~\eqref{eq:DSF_RBP} and \eqref{eq:JtoCt}, with the parameters obtained in references~\cite{MG23_primo,MG23_secundo}.}\label{Fig:CF110-1.3}
\end{figure}


\begin{figure}[h!]\centering
{\includegraphics[width=1.\textwidth]{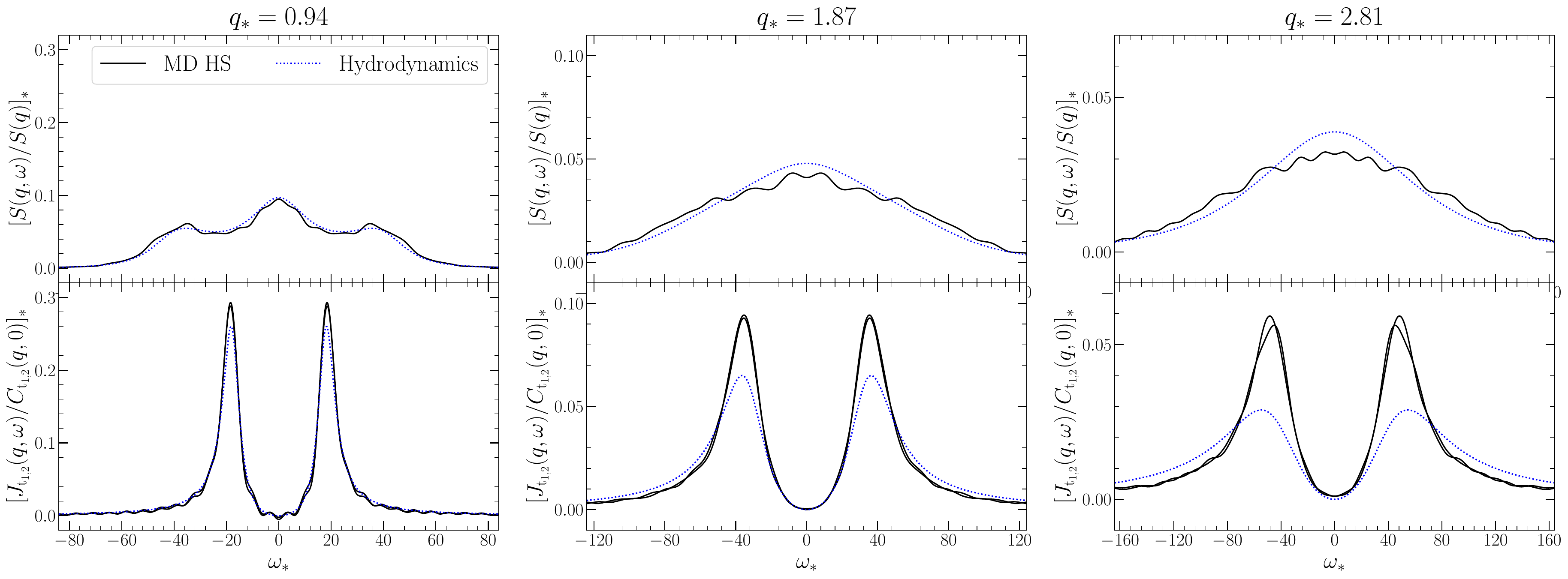}}
\caption[] {Normalized dynamic structure factor $S(q,\omega)$ and  spectral functions of transverse momentum density fluctuations $J_{{\rm t}_{1,2}}(q,\omega)$ versus the frequency at density $n_*=1.3$ for a solid of $N=2048$ hard spheres for ${\bf q}$ in the direction  $[111]$ with magnitudes $q_*\in\{0.94, 1.87, 2.81\}$. The solid lines correspond to the spectral functions obtained from molecular dynamics simulations of hard spheres. The dotted lines correspond to the analytical expressions predicted by hydrodynamics in equations~\eqref{eq:DSF_RBP} and \eqref{eq:JtoCt}, with the parameters obtained in references~\cite{MG23_primo,MG23_secundo}.}\label{Fig:CF111-1.3}
\end{figure}


\begin{sidewaysfigure}[t!] \centering
    \begin{minipage}[t]{0.3\textwidth}
        \centering
        \includegraphics[width=1.\textwidth,valign=t]{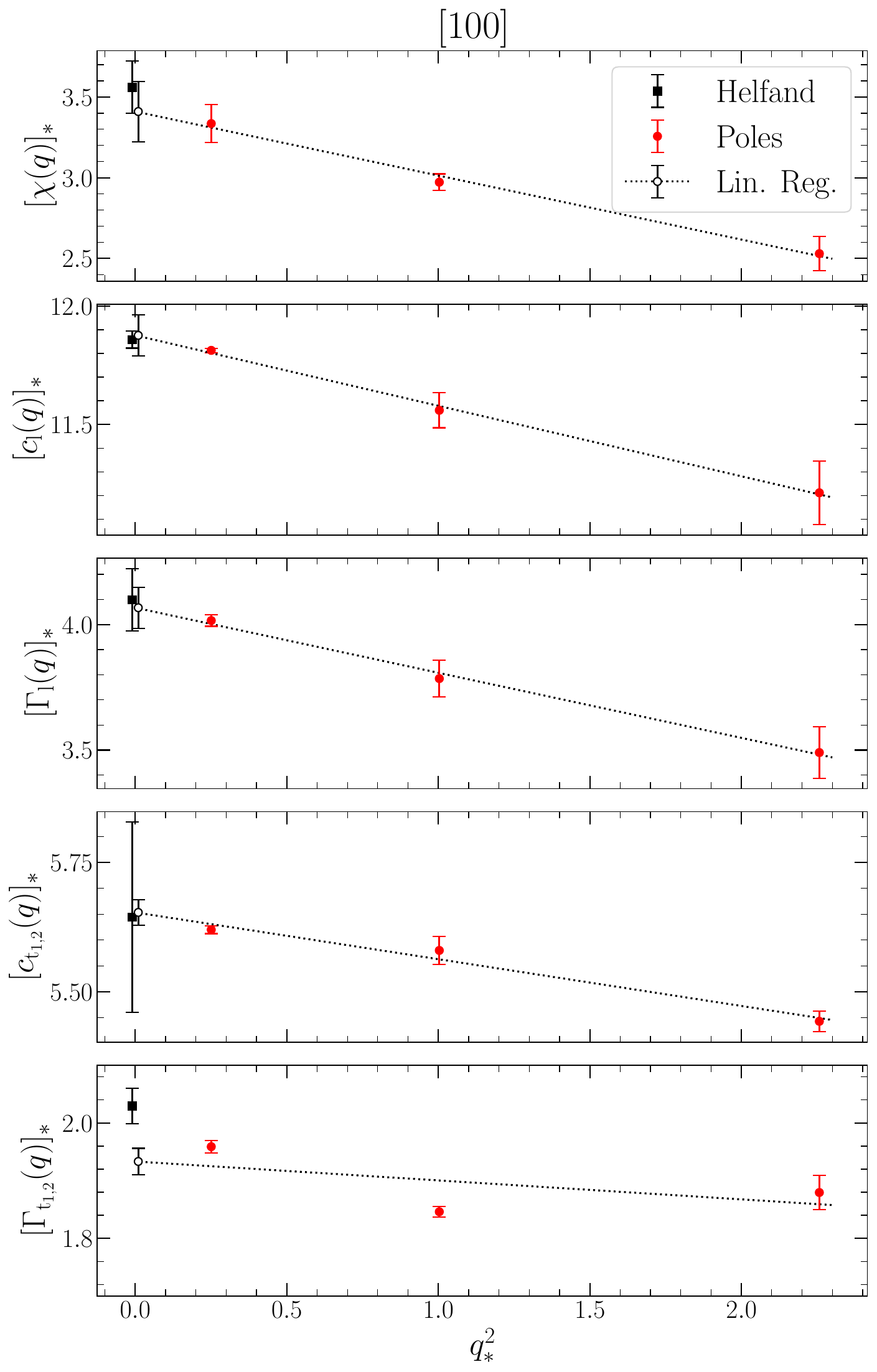} 
    \end{minipage}\hfill
        \begin{minipage}[t]{0.3\textwidth}
        \centering
        \includegraphics[width=1.\textwidth,valign=t]{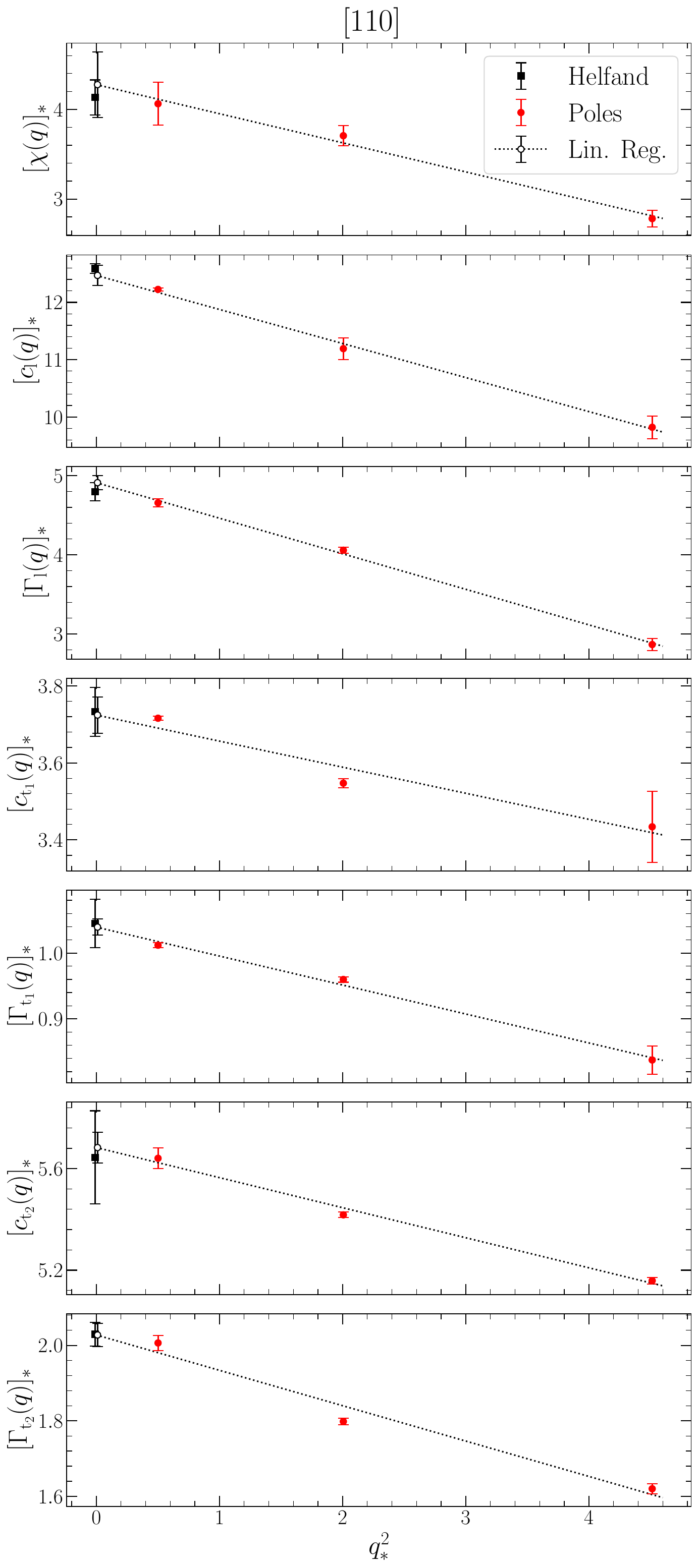} 
    \end{minipage}\hfill
        \begin{minipage}[t]{0.3\textwidth}
        \centering
        \includegraphics[width=1.\textwidth,valign=t]{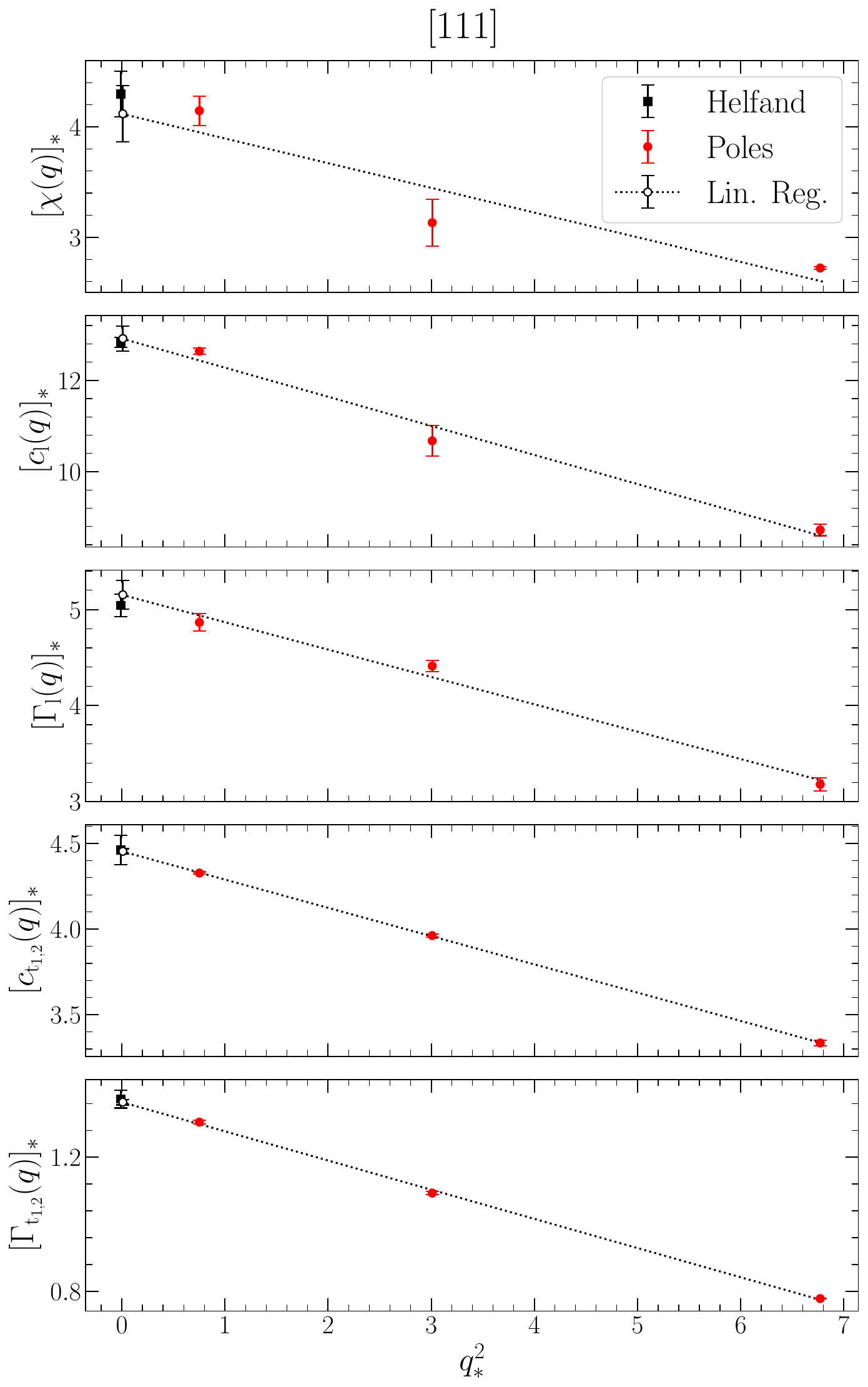} 
    \end{minipage}
\caption[] {Extrapolation to $q=0$ for the speeds of longitudinal and transverse sound waves and  the diffusivities obtained from the poles of the spectral functions $S(q,\omega)$ and $J_{{\rm t}_{1,2}}(q,\omega)$ with ${\bf q}$ along the directions $[100]$, $[110]$, and $[111]$ at a density $n_*=1.037$ for a solid of $N=2048$ hard spheres. The dotted line corresponds to the linear least square  regression. The value at $q=0$ shown with a square is obtained using the Helfand moments~\cite{MG23_primo}.}\label{Fig:LR_rho1.037}
\end{sidewaysfigure}


\begin{sidewaysfigure}[t!] \centering
    \begin{minipage}[t]{0.3\textwidth}
        \centering
        \includegraphics[width=1.\textwidth,valign=t]{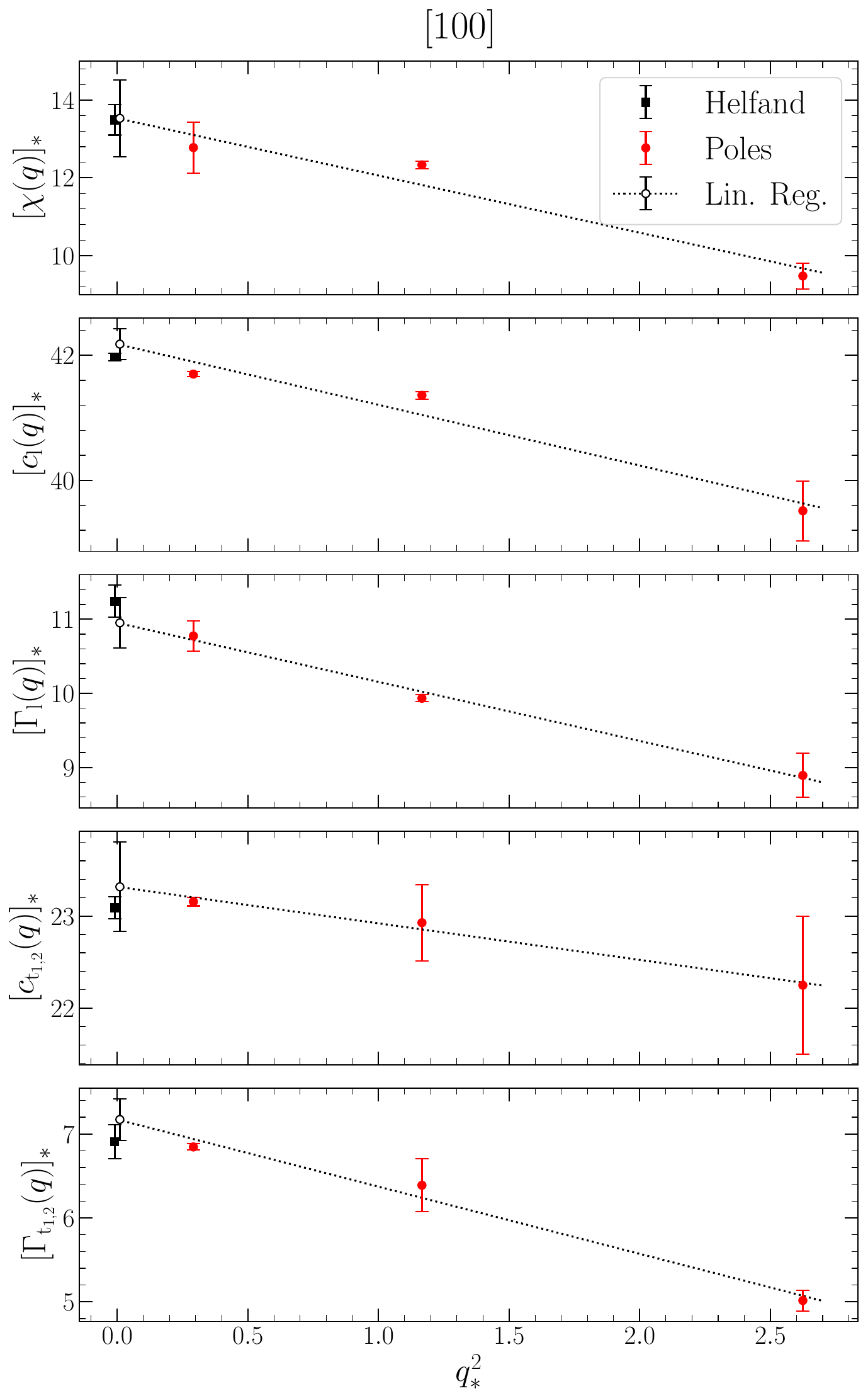} 
    \end{minipage}\hfill
        \begin{minipage}[t]{0.3\textwidth}
        \centering
        \includegraphics[width=1.\textwidth,valign=t]{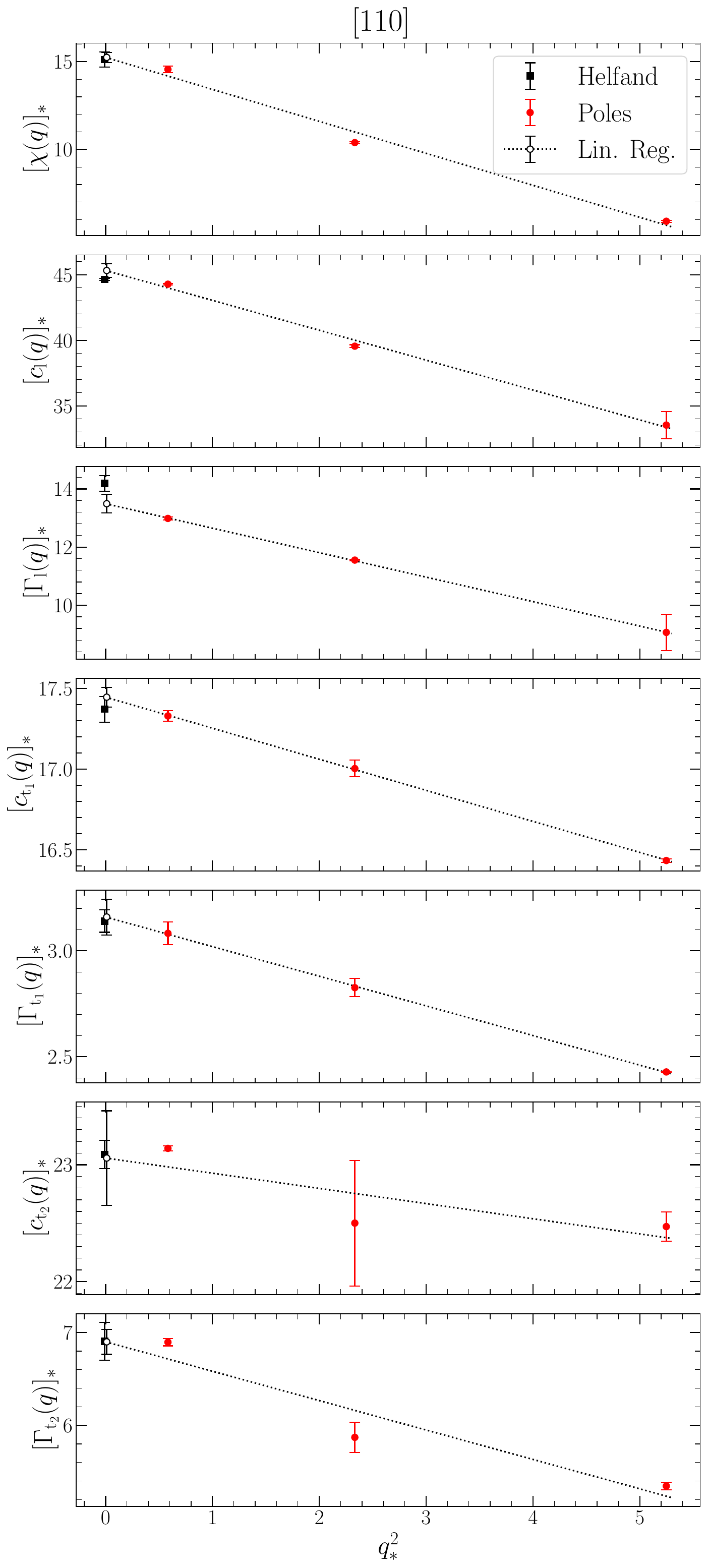} 
    \end{minipage}\hfill
        \begin{minipage}[t]{0.3\textwidth}
        \centering
        \includegraphics[width=1.\textwidth,valign=t]{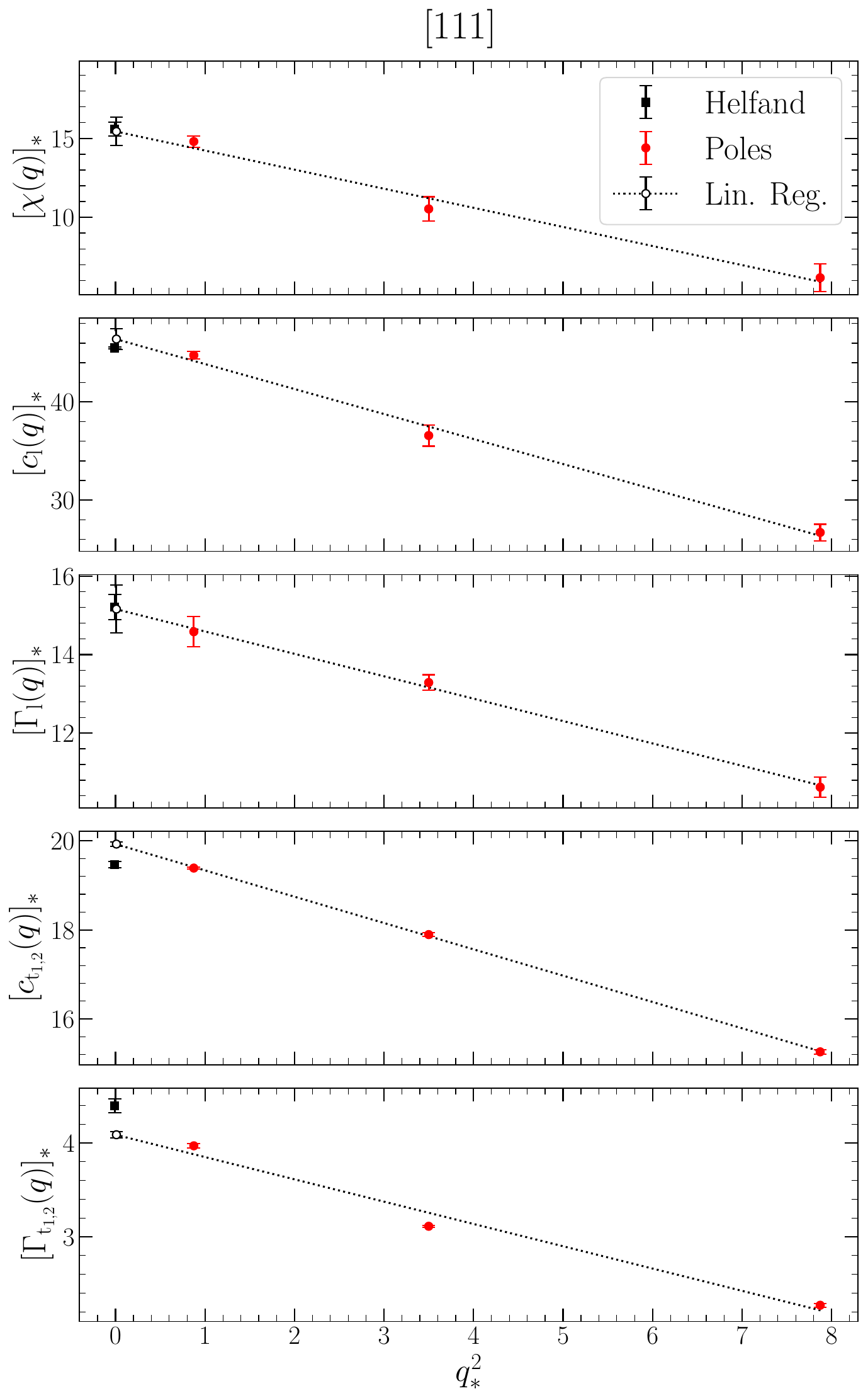} 
    \end{minipage}
\caption[] {Extrapolation to $q=0$ for  the speed of longitudinal and transverse sound waves  and the diffusivities obtained from the poles of the spectral functions $S(q,\omega)$ and $J_{{\rm t}_{1,2}}$ with ${\bf q}$ along the directions $[100]$, $[110]$, and $[111]$ at a density $n_*=1.3$ for a solid of $N=2048$ hard spheres. The dotted line corresponds to the linear least square  regression. The value at $q=0$ shown with a square is obtained using the Helfand moments~\cite{MG23_primo}.}\label{Fig:LR_rho1.3}
\end{sidewaysfigure}


\begin{figure}[t!] \centering
    \begin{minipage}[t]{0.3\textwidth}
        \centering
        \includegraphics[width=1.\textwidth,valign=t]{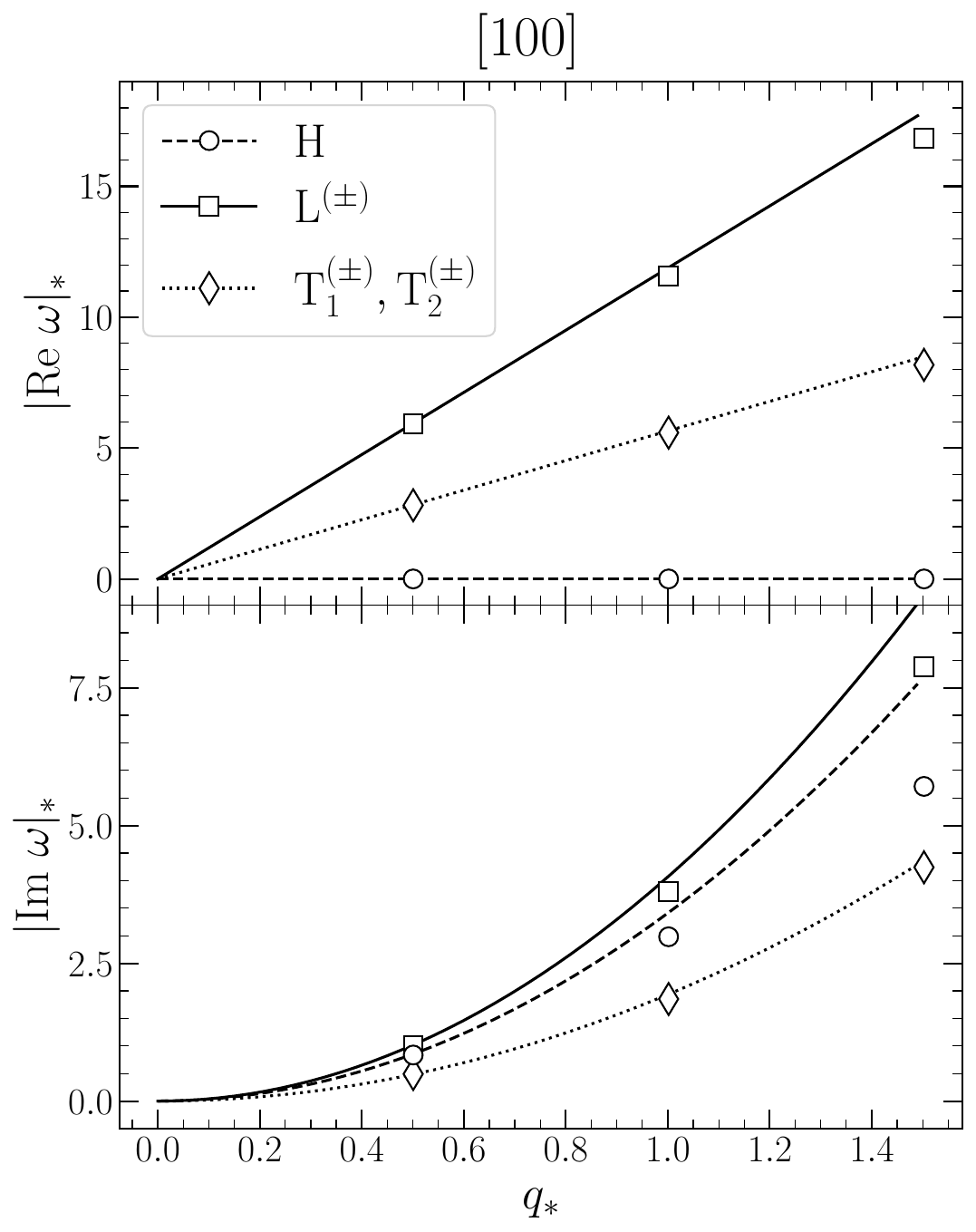} 
    \end{minipage}\hfill
        \begin{minipage}[t]{0.3\textwidth}
        \centering
        \includegraphics[width=1.\textwidth,valign=t]{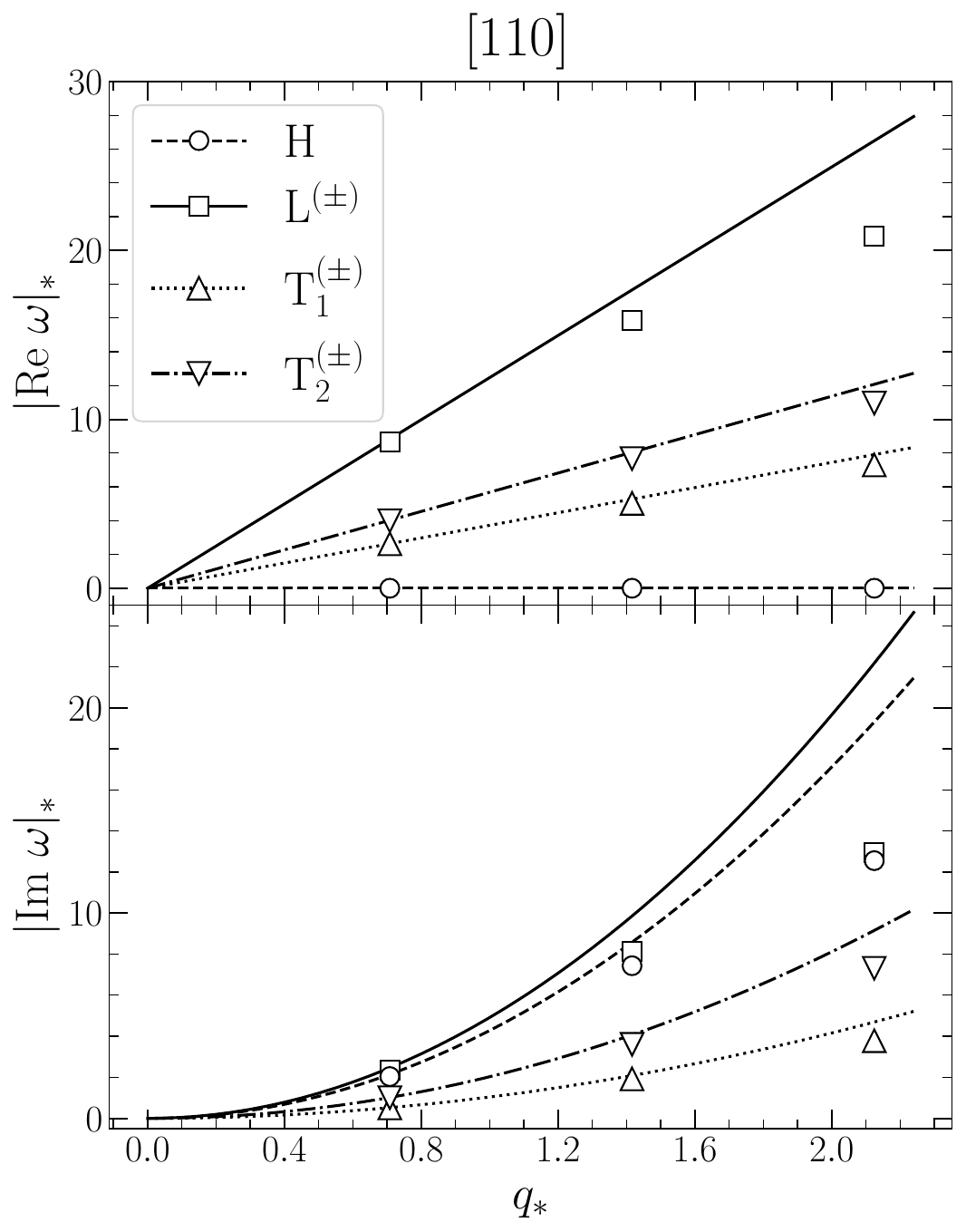} 
    \end{minipage}\hfill
        \begin{minipage}[t]{0.3\textwidth}
        \centering
        \includegraphics[width=1.\textwidth,valign=t]{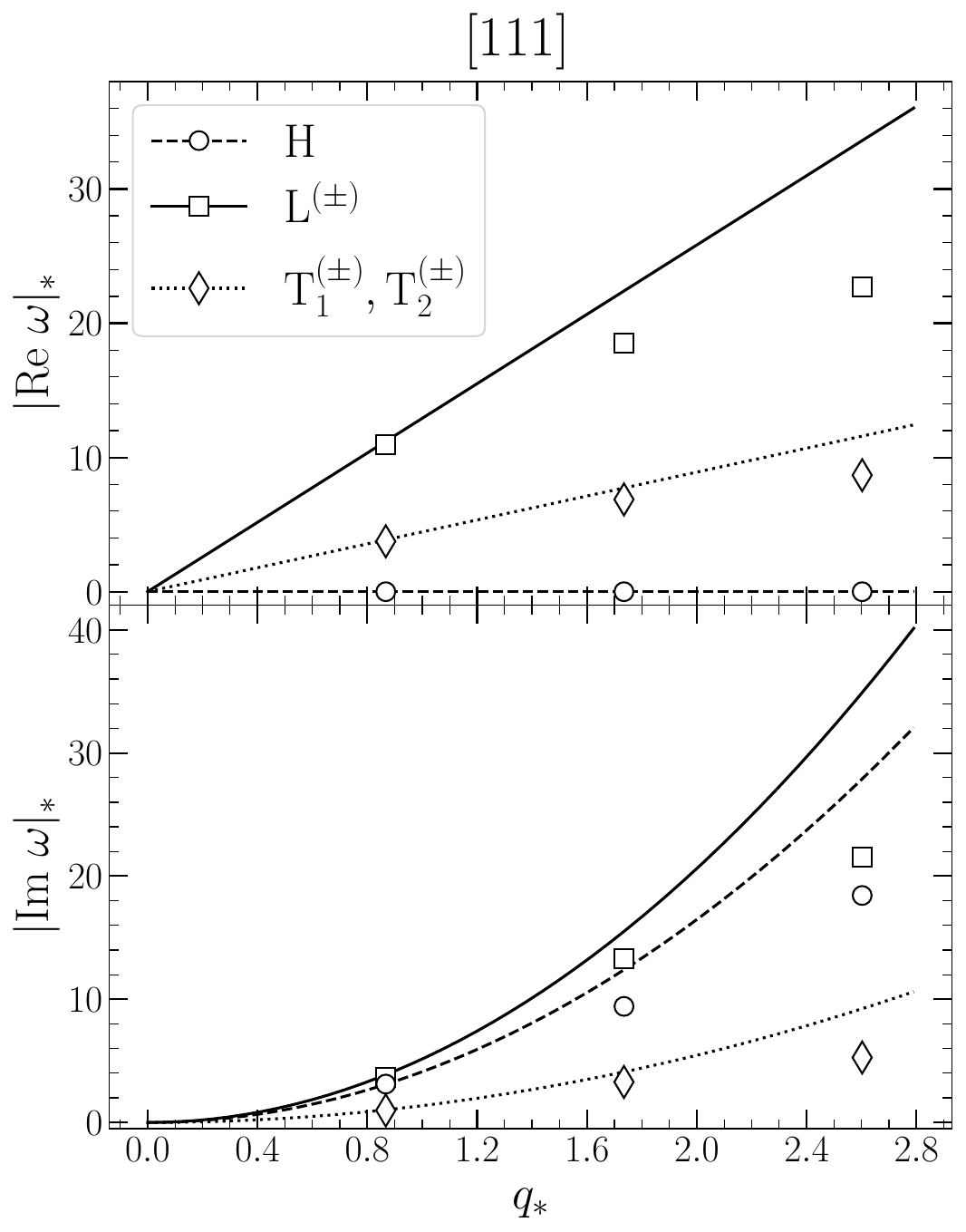} 
    \end{minipage}
\caption[] {Dispersion relations with ${\bf q}$ along the  directions [100], [110], and  [111], at a density $n_*=1.037$ for a solid of $N=2048$ hard spheres. The symbols correspond to the values at finite $q$ computed from the location of the poles and given in tables~\ref{Tab:CC1.037_100},~\ref{Tab:CC1.037_110}, and~\ref{Tab:CC1.037_111}. The lines show the dispersion relations obtained from the extrapolated values of the poles at $q=0$. H denotes the heat mode, ${\rm L}^{\pm}$ the two longitudinal sound modes, and ${\rm T}_{1,2}^{\pm}$ the four transverse sound modes.}\label{Fig:DR_1d037}
\end{figure}


\begin{figure}[t!] \centering
    \begin{minipage}[t]{0.3\textwidth}
        \centering
        \includegraphics[width=1.\textwidth,valign=t]{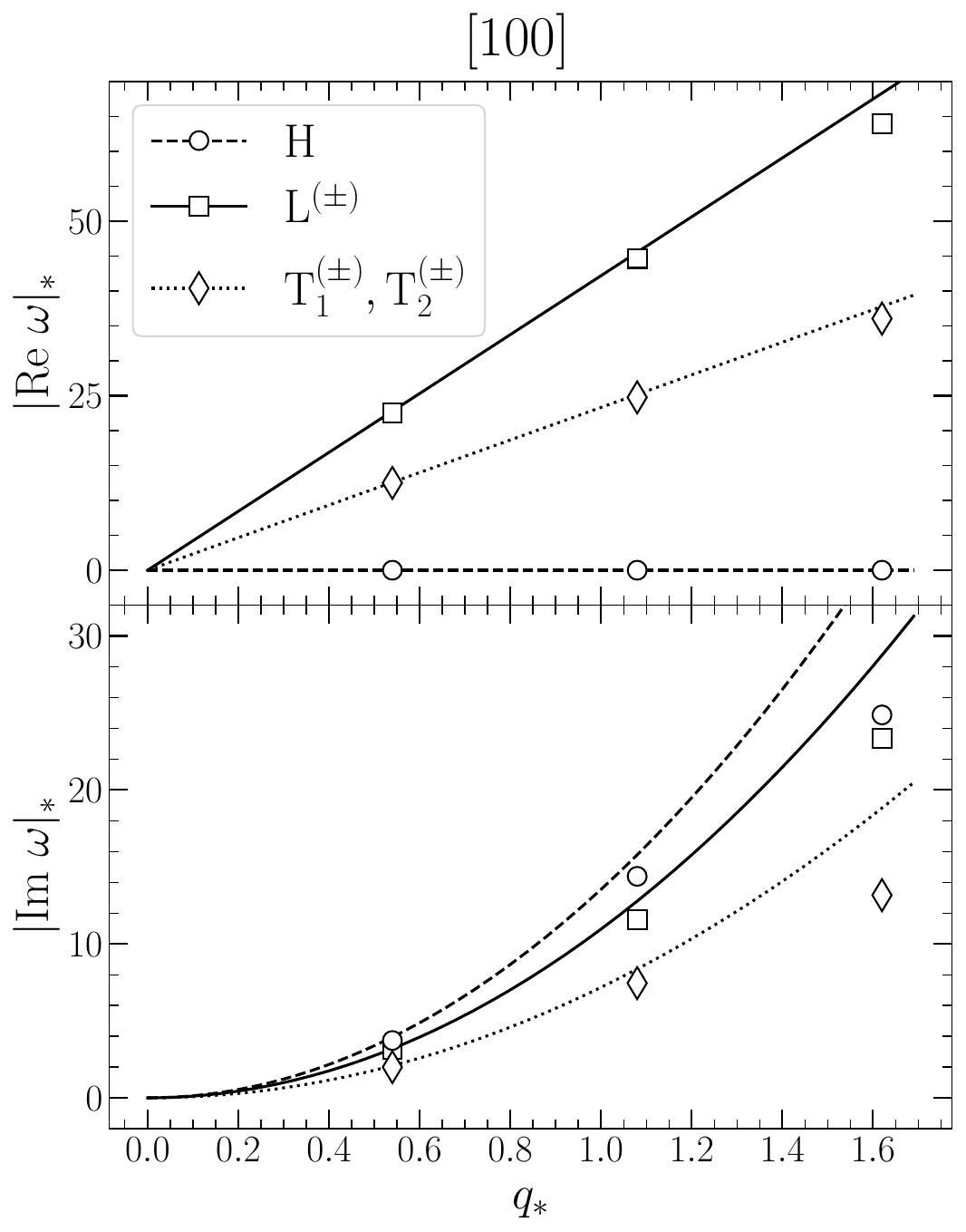} 
    \end{minipage}\hfill
        \begin{minipage}[t]{0.3\textwidth}
        \centering
        \includegraphics[width=1.\textwidth,valign=t]{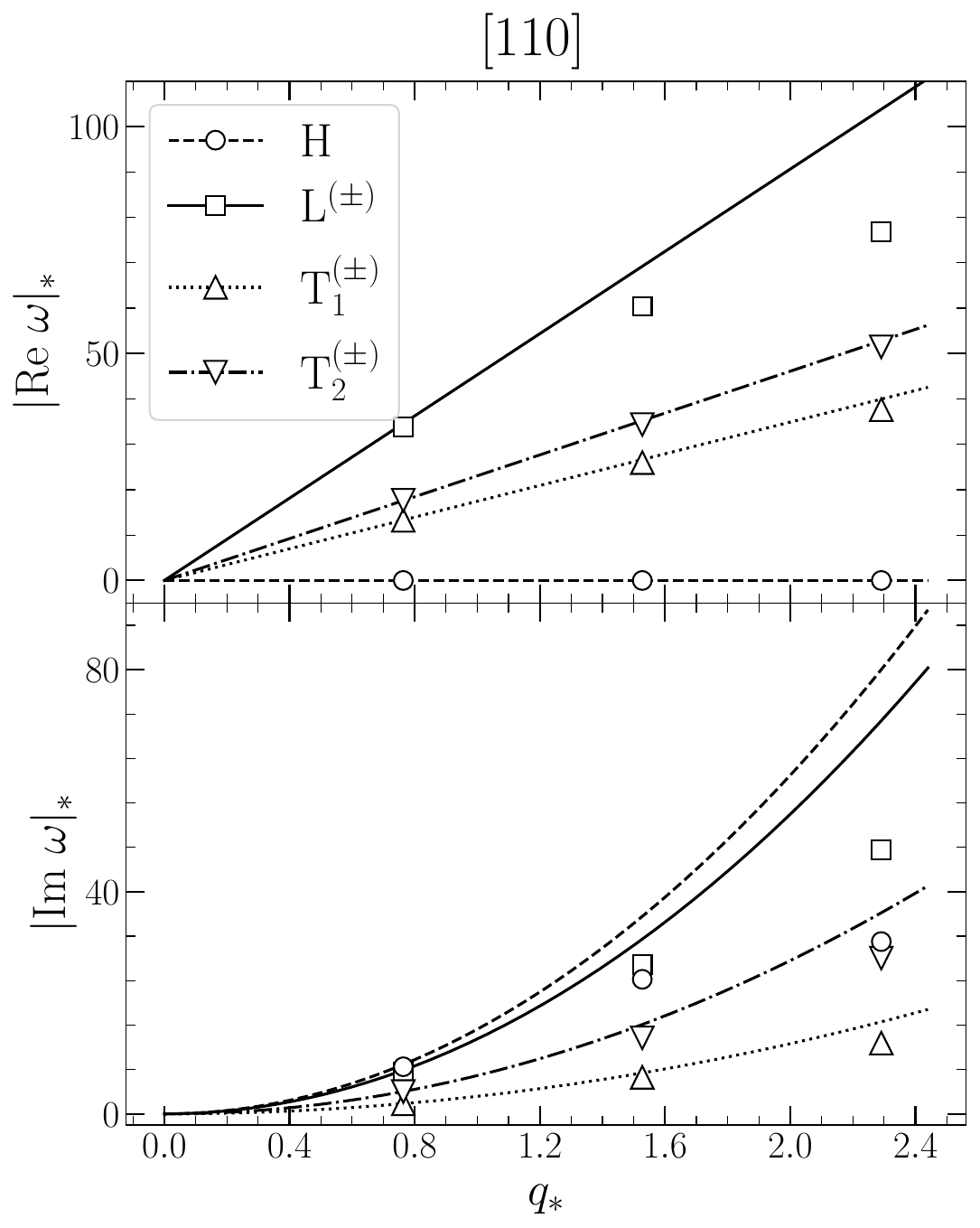} 
    \end{minipage}\hfill
        \begin{minipage}[t]{0.3\textwidth}
        \centering
        \includegraphics[width=1.\textwidth,valign=t]{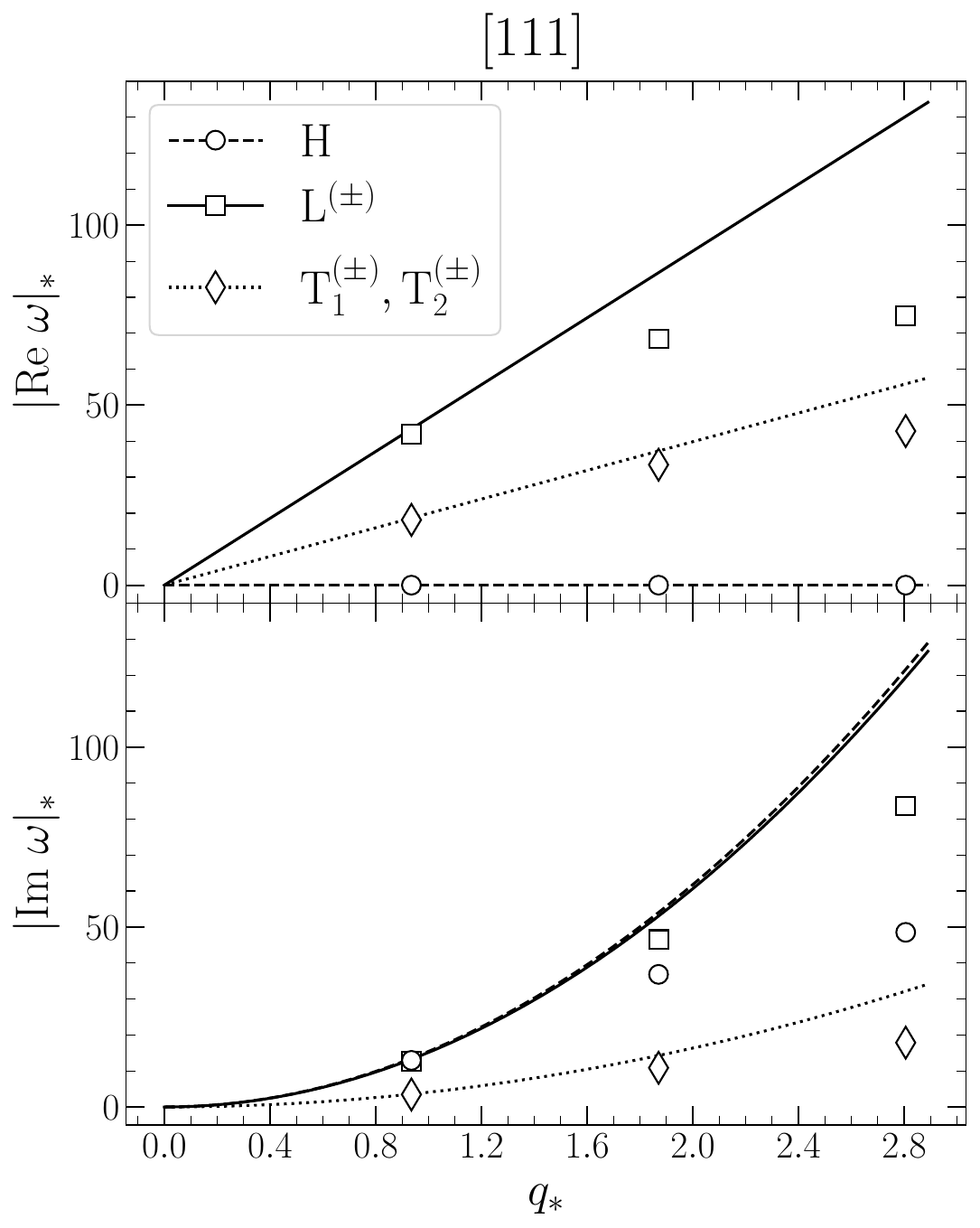} 
    \end{minipage}
\caption[] {Dispersion relations with ${\bf q}$ along the  directions [100], [110], and  [111], at a density $n_*=1.3$ for a solid of $N=2048$ hard spheres. The symbols correspond to the values at finite $q$ computed from the location of the poles and given in tables~\ref{Tab:CC1.3_100},~\ref{Tab:CC1.3_110}, and~\ref{Tab:CC1.3_111}. The lines show the dispersion relations obtained from the extrapolated values of the poles at $q=0$. H denotes the heat mode, ${\rm L}^{\pm}$ the two longitudinal sound modes, and ${\rm T}_{1,2}^{\pm}$ the four transverse sound modes.}\label{Fig:DR_1d3}
\end{figure}

\end{document}